\documentclass[superscriptaddress,amsmath,amssymb,aps,prb,notitlepage,twocolumn,floatfix]{revtex4-2}

\usepackage[utf8]{inputenc}
\usepackage{listings}
\usepackage{enumerate}
\usepackage{latexsym}
\usepackage{braket}
\usepackage[version=4]{mhchem}
\usepackage{graphicx}
\usepackage[colorlinks=True,linkcolor=red,citecolor=blue,urlcolor=blue]{hyperref}
\usepackage{blkarray}
\usepackage{array}
\usepackage{color}
\usepackage[normalem]{ulem}
\usepackage{wasysym}
\usepackage{pdfpages}
\makeatletter
\AtBeginDocument{\let\LS@rot\@undefined}
\makeatother

\newcommand{\y}{\ce{Y3Cu9(OH)19Cl8}}
\newcommand{\yd}{\ce{Y3Cu9(OD)19Cl8}}
\newcommand{\yperfect}{\ce{YCu3(OH)6Cl3}}
\newcommand{\SX}{$S_{\text{XRD}}$}
\newcommand{\SN}{$S_{\text{ND}}$}

\date{\today}

\begin{document}

%\title{Distorted kagome antiferromagnet: Phase diagram and application to Y-kapellasite}
\title{Phase diagram of a distorted kagome antiferromagnet and application to Y-kapellasite}

\author{Max Hering}
\email[]{max.hering@fu-berlin.de}
\affiliation{Helmholtz-Zentrum Berlin f\"{u}r Materialien und Energie, Hahn-Meitner Platz 1, 14109 Berlin, Germany}
\affiliation{Dahlem Center for Complex Quantum Systems and Fachbereich Physik, Freie Universit\"at Berlin, 14195 Berlin, Germany}
\author{Francesco Ferrari}
\email[]{ferrari@itp.uni-frankfurt.de}
\affiliation{Institute for Theoretical Physics, Goethe University Frankfurt,
Max-von-Laue-Straße 1, 60438 Frankfurt am Main, Germany}
\author{Aleksandar Razpopov}
\affiliation{Institute for Theoretical Physics, Goethe University Frankfurt,
Max-von-Laue-Straße 1, 60438 Frankfurt am Main, Germany}
\author{Igor I. Mazin}
\affiliation{Department of Physics and Astronomy, and Quantum Science and
Engineering Center, George Mason University, Fairfax, VA 22030, USA}
\author{Roser Valent\'\i}
\affiliation{Institute for Theoretical Physics, Goethe University Frankfurt,
Max-von-Laue-Straße 1, 60438 Frankfurt am Main, Germany}
\author{Harald O. Jeschke}
\affiliation{Research Institute for Interdisciplinary Science, Okayama University, Okayama 700-8530, Japan}
\author{Johannes Reuther}
\affiliation{Helmholtz-Zentrum Berlin f\"{u}r Materialien und Energie, Hahn-Meitner Platz 1, 14109 Berlin, Germany}
\affiliation{Dahlem Center for Complex Quantum Systems and Fachbereich Physik, Freie Universit\"at Berlin, 14195 Berlin, Germany}

\begin{abstract}
We investigate the magnetism of a previously  unexplored distorted spin-1/2  kagome model consisting of three symmetry-inequivalent  nearest-neighbor antiferromagnetic Heisenberg couplings $J_{\varhexagon}$, $J$ and $J'$, and  uncover a rich ground state phase diagram even at the classical level. Using analytical arguments and numerical techniques we identify a collinear $\vec{Q} = 0$ magnetic phase, two unusual non-collinear coplanar $\vec{Q} = (1/3,1/3)$ phases and a classical spin liquid phase with a degenerate manifold of non-coplanar ground states, resembling the jammed spin liquid phase found in the context of a bond-disordered kagome antiferromagnet. We further show with density functional theory calculations that the recently synthesized Y-kapellasite {\y} is a realization of this model and predict its ground state to lie in the region of $\vec{Q} = (1/3,1/3)$ order, which remains stable even after inclusion of quantum fluctuation effects within variational Monte Carlo and pseudofermion functional renormalization group.  The presented model opens a new direction in the study of kagome antiferromagnets.
\end{abstract}

\maketitle

\vspace{0.2cm}{\bf\large Introduction}

The kagome lattice is arguably one of the most important two-dimensional (2D) lattices for the study of magnetic frustration. It is characterized by a complex phase diagram including magnetically ordered regimes and proposed quantum spin liquid phases~\cite{Bieri2016}, has rich magnetization dynamics~\cite{Nishimoto2013}, and supports some of the best studied quantum spin liquid candidates like herbertsmithite ZnCu$_3$(OH)$_6$Cl$_2$~\cite{Mendels2010,han2012,Norman2016,mendels2020}. In a more technical context, the study of the antiferromagnetic Heisenberg model on the kagome lattice has been a fertile ground for the development and benchmarking of theoretical methods. Notably, the competition between density matrix renormalization group (DMRG)~\cite{Depenbrock2012,Jiang2012,He2017}, variational Monte Carlo (VMC)~\cite{Iqbal2013,Iqbal2015b}, and tensor networks (TNs)~\cite{Liao2017} type methods, with the aim of resolving the nature of the spin liquids supported by the kagome lattice, has been a fervent area of research for many years.

All these intense research activities have mainly focused on the ideal kagome structure. In contrast, distortions of this lattice have been studied much less, even though they are realized in some magnetic compounds, and their physical phenomenology may be even richer than for the standard kagome lattice. In some cases, like volborthite Cu$_3$V$_2$O$_7$(OH)$_2\cdot 2$H$_2$O~\cite{Yoshida2012,Janson2016}, the distortion leads to a new 2D lattice which is still highly frustrated and possibly has a spin liquid ground state~\cite{Watanabe2016}. In \ce{Rb2Cu3SnF12}, the deformed kagome lattice leads to a pinwheel valence bond solid~\cite{Matan2010}. Other kinds of distortions lower the rotational symmetry of the lattice and lead to kagome strips~\cite{Goto2016,Jeschke2019}. Even the low temperature structure of herbertsmithite bears some signatures of distortion~\cite{Zorko2017,Laurita2019,li2020}.

The focus of the present work lies on an unusual and previously unexplored distortion of the kagome lattice which is realized in the recently synthesized variant of herbertsmithite, namely {\y}. The distorted lattice structure consists of three symmetry-inequivalent nearest-neighbor kagome bonds forming a nine-site unit cell. Analyzing the corresponding Heisenberg model as a function of its two coupling ratios, using analytical arguments and numerical techniques, we find a surprisingly rich ground state phase diagram, even at the classical level. A first notable observation is that large parts of the phase diagram represent an unusual coplanar spin state with a commensurate magnetic wave vector $\vec{Q}=(1/3,1/3)$. This type of ordered state requires a $27$ atom magnetic unit cell. Furthermore, in an extended regime around the standard undistorted kagome lattice, an even more complex classical spin liquid phase is identified, which cannot be characterized by any specific wave vector. It bears similarities with the well-known classical spin liquid on the undistorted kagome lattice in the sense that its low energy states follow from a set of spin constraints for each triangle~\cite{Moessner1998}. In contrast, however, the ground state in the distorted case is found to be generally non-coplanar and, hence, resembles the jammed spin liquid investigated in Ref.~\cite{Moessner2017}. 

After discussing in detail the general magnetic phenomena of this distorted kagome lattice, the second focus of this paper is on the specific case of {\y} where this model is likely realized. This material was discovered in an attempt of electron doping the Cu $3d$ states in herbertsmithite with the aim of placing the Fermi level at the symmetry protected Dirac crossing of the Cu $d$-bands in the kagome lattice~\cite{Mazin2014,guterding2016kagome,Puphal2017}. In herbertsmithite-type copper hydroxy halides, however, the larger charge provided by Y$^{3+}$ compared to Zn$^{2+}$ is always compensated by the incorporation of additional hydroxy or halide ions, preserving the antiferromagnetic insulator nature of the Cu$^{2+}$ layers. 
Even if charge doping remains elusive in these systems, the newly discovered by-product in form of  {\y} appears to be of great interest in and of itself.

In {\y}, the Y$^{3+}$ ions are placed in the center of the hexagon of the kagome lattice, making it a material which is structurally similar to kapellasite~\cite{Fak2012,Bieri2015,Iqbal2015}, haydeeite~\cite{Boldrin2015} or centennialite~\cite{Doki2018,Iida2020}. We therefore name the system Y-kapellasite. Note that there is a closely related compound {\yperfect} with ideal kagome lattice but disordered Y positions~\cite{Sun2016}. The latter orders at $T_{\rm N}=15$~K in a $\vec{Q}=0$ structure with negative spin chirality~\cite{Zorko2019a,Zorko2019b} which has been attributed to a strong Dzyaloshinskii-Moriya (DM) interaction~\cite{Arh2020}. In contrast, Y-kapellasite remains dynamical down to much lower temperatures than {\yperfect}~\cite{Barthelemy2019}; it has a broad feature at $T=2$\,K in the specific heat but muon spin resonance ($\mu$SR) on powder samples seems to indicate the absence of any static magnetic order, although disorder effects may play a role. Recently, coexistence of magnetic order and persistent spin dynamics has been suggested for different samples of Y-kapellasite~\cite{Sun2021}.

By extracting the Heisenberg Hamiltonian of Y-kapellasite using total energy mapping from density functional theory (DFT) calculations, we find that the three couplings on the symmetry-inequivalent nearest-neighbor kagome bonds dominate, with negligible longer range interactions. We may, hence, place Y-kapellasite in the region of $\vec{Q}=(1/3,1/3)$ order in the classical ground state phase diagram obtained here. Investigating the corresponding spin-1/2 model within variational Monte Carlo (VMC) and pseudofermion functional renormalization group (PFFRG) we argue that quantum fluctuations are not sufficiently strong to suppress the long-range magnetic order. Accordingly, our semiclassical spin-wave analysis provides a realistic approximation of the system's excitation spectrum which will be useful for comparison with future experimental data. 

\vspace{0.2cm}{\bf\large Results}

\begin{figure*}
	\centering
	\includegraphics[width=1.7\columnwidth]{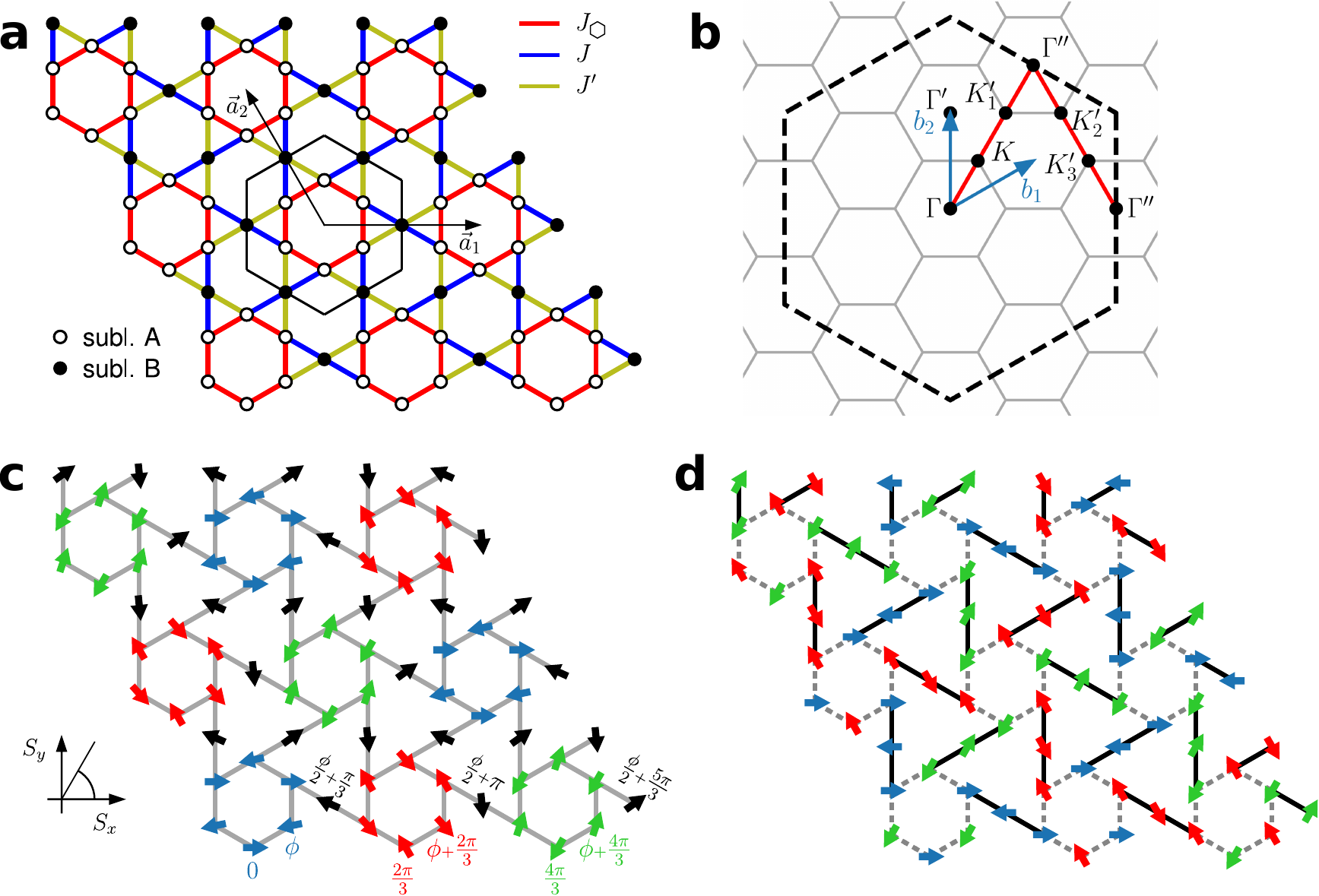}
	\caption{{\bf Distorted kagome lattice and classical $\mathbf{\vec{Q}=(1/3,1/3)}$ magnetic order.} {\bf a} Schematic illustration of the three exchange couplings characterizing the effective Heisenberg Hamiltonian for Y-kapellasite ($J_{\varhexagon}$, $J$ and $J'$) shown in red, blue and green, respectively. The presence of three different couplings breaks the translational symmetry of the kagome lattice and leads to a decorated triangular lattice with an enlarged unit cell of $9$ sites, here represented by the black hexagon ({Wigner-Seitz cell}). The Hamiltonian of the system is periodic under translations along the Bravais vectors $\vec{a}_1$ and $\vec{a}_2$ and the sites within the unit cell are divided into sublattices $A$ (hollow symbols) and $B$ (solid symbols). Due to the different values of the three exchange terms, the $D_6$ point group symmetry of the kagome lattice is broken down to $C_6$. {\bf b} Pictorial view of the reciprocal space. The blue arrows represent the unit vectors of the reciprocal space ($\vec{b}_1$ and $\vec{b}_2$). The gray hexagons tiling the reciprocal space depict the first Brillouin zone of the lattice, while the black dashed hexagon delimits the so-called extended Brillouin zone. Some of the high symmetry points are marked with black dots. Finally, red lines represent the path along which the magnon dispersion is plotted in Fig.~\ref{fig:spinwaves_1}. {\bf c} Classical $\vec{Q}=(1/3,1/3)$ magnetic order for ${J>J'}$ (red region of Fig.~\ref{fig:Classical-PD}). The orientations of the spins are fully specified by the angle $\phi$ between neighboring spins in the $J_{\varhexagon}$-hexagons, as outlined in the main text (here we take the value of $\phi$ for the case $J'=0$ and $J_{\varhexagon}=J$). In this figure, the spins are arranged in the $xy$-plane and their orientation is represented by the angle with respect to the $S_x$ axis. The red, blue and green colors of the spins of sublattice $A$ help visualizing the ${\vec{Q}=(1/3,1/3)}$ pattern. 
	{\bf d} The classical $\vec{Q}=(1/3,1/3)$ magnetic order of
	Fig.~\ref{fig:ninja_lat}\,c in the $J \gg J_{\varhexagon}$ limit ($J'=0$). The spins form antiferromagnetic trimers along the $J$-bonds (depicted in black). The trimers are arranged in an effective kagome lattice structure and their orientations, highlighted by the three different colors, follow the  $\sqrt{3}\times \sqrt{3}$ pattern~\cite{reimers1993}. 
	\label{fig:ninja_lat}}
\end{figure*}

{\bf Spin Hamiltonian.} The model investigated in this work is a variant of the standard nearest-neighbor kagome Heisenberg model, but with three distinct nearest-neighbor couplings, which we call $J$, $J_{\varhexagon}$, and $J'$ [see Fig.~\ref{fig:ninja_lat}\,a]. We will later argue that this model  approximates well the microscopic interactions in Y-kapellasite. The Heisenberg Hamiltonian can be written as
\begin{equation}\label{eq:heis_ham}
\mathcal{H}=\sum_{\langle i,j \rangle} J_{ i j } \vec{S}_i \cdot \vec{S}_j\;,
\end{equation}
where $\vec{S}_i$ are the spin degrees of freedom (which, below, are either chosen as spin-1/2 operators or as classical normalized vectors) and $J_{ i j }$ is given by $J$, $J_{\varhexagon}$ or $J'$, depending on the bond. All these couplings are assumed to be positive (antiferromagnetic). It is clear that $J=J_{\varhexagon}=J'$ leads back to the standard undistorted nearest-neighbor kagome model. As a consequence of the broken translational symmetry of the kagome lattice the system's periodic structure is described by a decorated triangular lattice with a unit cell of nine sites [see Fig.~\ref{fig:ninja_lat}\,a]. We can distinguish two inequivalent sets of sites inside the unit cell, which are not connected by point group symmetries and form two distinct sublattices: sublattice $A$ is made of the six sites connected by $J_{\varhexagon}$ [the vertices of the red hexagons of Fig.~\ref{fig:ninja_lat}\,a]; sublattice $B$ is made of the  remaining three sites. Also note that the model is invariant under exchanging $J$ and $J'$ followed by a reflection with respect to the $\vec{a}_1$ axis.

\begin{figure}
	\centering
	\includegraphics[width=\columnwidth]{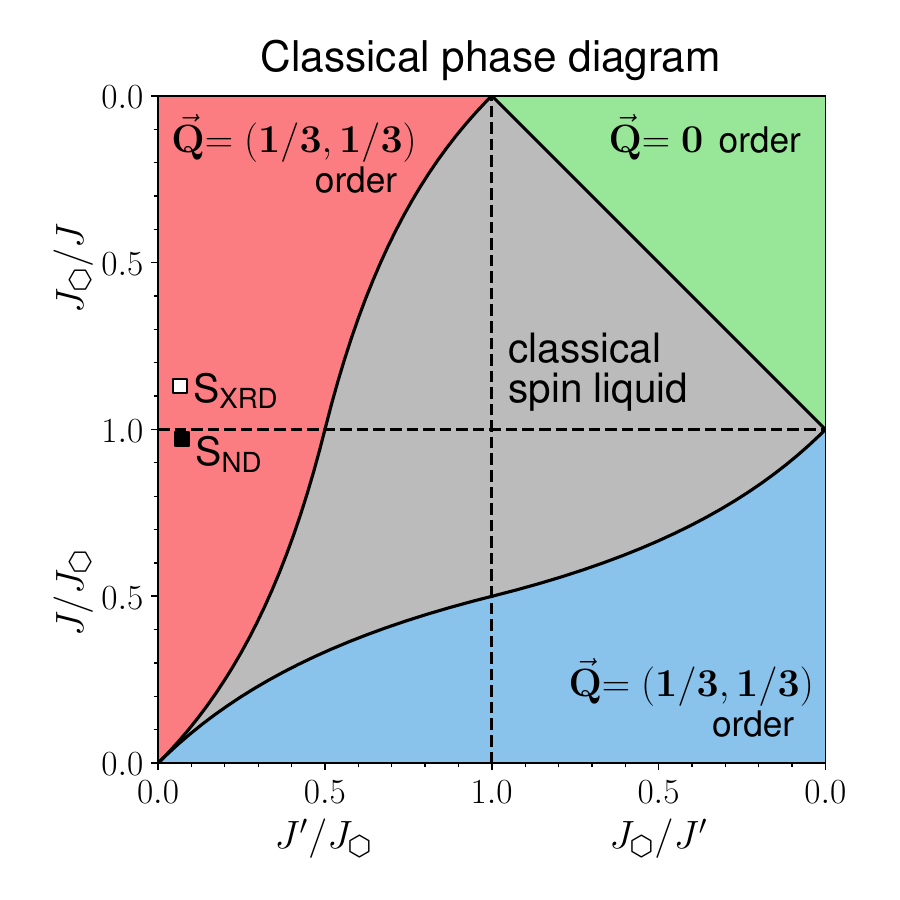}
	\caption{{\bf Classical phase diagram of the distorted kagome model.}
	  We note that the phase diagram is symmetric under the exchange of the axes (i.e., $J\leftrightarrow J'$), as a consequence of the symmetry of the Hamiltonian. The $\vec{Q}=(1/3,1/3)$ magnetic order of the red region is depicted in Fig.~\ref{fig:ninja_lat}\,c, and related to the $\vec{Q}=(1/3,1/3)$ order of the blue region by a mirror reflection with respect to $\vec{a}_1$. Inside the gray area, the system features a classical spin-liquid phase with degenerate non-coplanar ground states, as discussed in the main text.%Sec.~\ref{subsection:Non_coplanar}.
          The $\vec{Q}=0$ magnetic order can be viewed as parallel spins on the same sublattice, while on different sublattices the spins are anti-aligned. We note that the axes change between the four quadrants of this plot. The empty and filled squares indicate two possible sets of couplings for Y-kapellasite.% discussed in Sec.~\ref{sec:abinitio}.
          \label{fig:Classical-PD}}
\end{figure}

%\label{sec:phasediagram}
{\bf Classical phase diagram.} In Fig.~\ref{fig:Classical-PD} we summarize the classical ground state phase diagram of the Heisenberg model on the distorted kagome lattice [Eq.~\eqref{eq:heis_ham}] as a function of the ratios $J/J_{\varhexagon}$ and $J'/J_{\varhexagon}$, which has been obtained by combining analytical arguments, iterative minimization and classical Monte Carlo calculations (see Methods). At the classical level, we observe (i) a collinear $\vec{Q}=0$ magnetic phase, (ii) two non-collinear coplanar magnetic phases, both labelled as $\vec{Q}=(1/3,1/3)$ order, separated by (iii) a classical spin-liquid phase with a degenerate manifold of non-coplanar ground states, which in the context of a bond-disordered kagome antiferromagnet was dubbed a ''jammed spin liquid'' phase~\cite{Moessner2017}. 

Even without any prior knowledge of the precise nature of the different phases, there exists a simple argument that determines the location of the phase boundaries. To this end, we employ the analytical procedure of Ref.~\cite{Moessner2017} where a bond disordered Heisenberg  model on the kagome lattice was studied. In the first step, we rewrite the Hamiltonian of Eq.~\eqref{eq:heis_ham} as
\begin{equation}
    \mathcal{H}=\frac{1}{2}\sum\limits_{\triangle}\big(\vec{L}_{\triangle}\big)^2+\text{const.}\,\label{ham_triangle}
\end{equation}
where the sum runs over all triangles formed by nearest-neighbor bonds of the kagome lattice (both up and down triangles are considered). We define 
\begin{equation}\label{eq:Ltriang}
    \vec{L}_{\triangle}=\sqrt{\frac{J_{ij}J_{ik}}{J_{jk}}}\vec{S}_{i}+\sqrt{\frac{J_{ji}J_{jk}}{J_{ik}}}\vec{S}_{j}+\sqrt{\frac{J_{ki}J_{kj}}{J_{ij}}}\vec{S}_{k} \,,
\end{equation}
where $i,\,j,\,k\in\triangle$ are the three sites forming a triangle. In our distorted model, all triangles of the lattice are formed by one $J_{\varhexagon}$, one $J$, and one $J'$ coupling [see Fig.~\ref{fig:ninja_lat}\,a]. Thus, by an appropriate choice of the $i,j,k$ labels of Eq.~\eqref{eq:Ltriang}, we can write
\begin{equation}
    \vec{L}_{\triangle}=\sqrt{JJ'/J_{\varhexagon}}\vec{S}_{i}+\sqrt{JJ_{\varhexagon}/J'}\vec{S}_{j}+\sqrt{J'J_{\varhexagon}/J}\vec{S}_{k} \,\label{def_l}
\end{equation}
for all triangles. From Eq.~(\ref{ham_triangle}) it immediately follows that any spin configuration that fulfills the condition $\vec{L}_{\triangle}=0 \ \ \forall \triangle$
is a ground state of the system. However, depending on the values of the couplings $J$, $J_{\varhexagon}$, and $J'$, it may occur that $\vec{L}_{\triangle}=0$ is impossible for any triangle when one term on the right hand side of Eq.~(\ref{def_l}) dominates so strongly that it cannot be compensated by the other two terms. 

Restricting to an isolated triangle, it is easy to show that $\vec{L}_{\triangle}=0$ can only be fulfilled if
\begin{subequations}\label{eq:NonCPCondition}
\begin{align}
    J/J_{\varhexagon}&\leq J'/(J_{\varhexagon}-J') \,,\quad J'\leq \text{min}(J,J_{\varhexagon}) \, , \\ J/J_{\varhexagon}&\geq J'/(J_{\varhexagon}+J') \,,\quad J\leq \text{min}(J',J_{\varhexagon}) \, ,\\ J/J_{\varhexagon}&\geq J'/(J'-J_{\varhexagon}) \,,\quad J_{\varhexagon} \leq \text{min}(J,J') \, .
\end{align}
\end{subequations}
These conditions define the phase boundaries in Fig.~\ref{fig:Classical-PD}. In the regions where an isolated triangle cannot satisfy $\vec{L}_{\triangle}=0$, the system realizes one of the aforementioned coplanar phases [$\vec{Q}=(1/3,1/3)$ order and $\vec{Q}=0$ order]. On the other hand, in the region where an isolated triangle can fulfill Eq.~(\ref{eq:NonCPCondition}) we observe a classical spin-liquid phase. We note that analogous phase boundaries characterize the classical phase diagram of the square-kagome antiferromagnet~\cite{Morita2018bis}.

{\bf Coplanar orders.} We start our discussion of the classical ground states with the coplanar phases where $\vec{L}_{\triangle}=0$ is necessarily violated. The rewritten Hamiltonian in Eq.~(\ref{ham_triangle}) still implies that these phases form in a way that minimizes $(\vec{L}_{\triangle})^2$. To simplify the investigation, we first restrict ourselves to the case $J'=0$ where the $\vec{Q}=(1/3,1/3)$ phase is realized. In the phase diagram of Fig.~\ref{fig:Classical-PD}, this corresponds to the leftmost vertical axis and it will turn out to provide a good approximation of the exchange couplings of Y-kapellasite determined by the \textit{ab initio} DFT calculations (marked with squares in the figure).

In the limit $J'=0$, the model consists of a lattice of hexagons, made of sublattice $A$ sites, which are connected to each other through the $J$-trimers involving sublattice $B$ sites [Fig.~\ref{fig:ninja_lat}\,a]. Note that the middle spin of each trimer is fixed in the direction opposite to the sum of the edge spins. The magnetic order realized along the $J'=0$ line is depicted schematically in Fig.~\ref{fig:ninja_lat}\,c.
 The spins are coplanar and form a periodic configuration with momentum $\vec{Q}=(1/3,1/3)$ (in units of the reciprocal lattice vectors $\vec{b}_1$ and $\vec{b}_2$). This momentum corresponds to the $K$ point of the Brillouin zone of the lattice (and the symmetry related points), see Fig.~\ref{fig:ninja_lat}\,b. Within a given unit cell, the spins of sublattice $A$ form an alternating pattern around the $J_{\varhexagon}$-hexagons: the spins on even and odd sites are ferromagnetically aligned along two different directions, which are rotated with respect to each other by an angle $\phi$. The orientations of the spins on the remaining sites (i.e., sublattice $B$), which are only two-coordinated, are uniquely determined by the value of the angle $\phi$. Thus, in the limit $J'=0$, we can express the classical energy per site of the $\vec{Q}=(1/3,1/3)$ order as a simple function of $\phi$:
\begin{equation}
E/N=\frac{2}{3}\left[J_{\varhexagon} \cos(\phi) + J \cos\left(\frac{\phi}{2}+\frac{\pi}{3} \right)  \right]
\end{equation}
Minimization yields optimal angles $\phi$ that go from $\phi=\pi$ in the strong hexagon limit $J \ll J_{\varhexagon}$ to $\phi=\frac{4\pi}{3}$ in the trimer limit $J \gg J_{\varhexagon}$ (see Fig.~\ref{fig:ninja_lat}\,d and Supplementary Note 2).

Going away from the $J'=0$ limit, the $\vec{Q}=(1/3,1/3)$ order extends for a finite region along the $J'/J_{\varhexagon}$ axis (red area in Fig.~\ref{fig:Classical-PD}), which is bounded by the onset of the classical spin-liquid phase. Within this region, the numerical minimization of the classical energy shows that the spin pattern is unchanged with respect to the one shown in Fig.~\ref{fig:ninja_lat}\,c and the orientation of the spins is still determined only by the angle $\phi$ between the spins of sublattice $A$. As already mentioned, the phase diagram is invariant under the exchange of $J$ and $J'$, and a corresponding $\vec{Q}=(1/3,1/3)$ order is observed also in the proximity of the $J=0$ limit (blue area in Fig.~\ref{fig:Classical-PD}). The two $\vec{Q}=(1/3,1/3)$ ordered phases, one for $J>J'$ and the other for $J<J'$, are transformed into each other by a mirror reflection with respect to $\vec{a}_1$. In the numerical calculations, the two phases can be distinguished by their spin susceptibility in momentum space, defined as
\begin{equation}\label{eq:SpinSusceptibility}
 \chi_{\vec{k}}=\frac{1}{N}\sum_{i,j} e^{i \vec{k} \cdot (\vec{r}_i - \vec{r}_j)} \left\langle \vec{S}_i \cdot \vec{S}_j \right \rangle \, ,
\end{equation}
where $\vec{r}_i$ is the position of site $i$ in the kagome lattice, $N$ is the total number of sites and the brackets $\langle \dots \rangle$ denote an appropriate average. In the $\vec{Q}=(1/3,1/3)$ phase at $J>J'$, $\chi_{\vec{k}}$ displays high intensity peaks at the $K_2'$ points, while in the ordered phase at $J<J'$, its maxima are located at the $K'_3$ points [see Fig.~\ref{fig:ninja_lat}\,b].
In the limit where $J_{\varhexagon}\ll J,J'$ the system is no longer frustrated since each coupled neighbor of an $A$ site is a $B$ site and vice versa. The ground state order is, hence, given by a simple collinear $\vec{Q}=0$ state where the two sublattices have opposite spin orientations. This regime is marked green in Fig.~\ref{fig:Classical-PD}.

\begin{figure*}
	\centering
    	\includegraphics[width=0.95\textwidth]{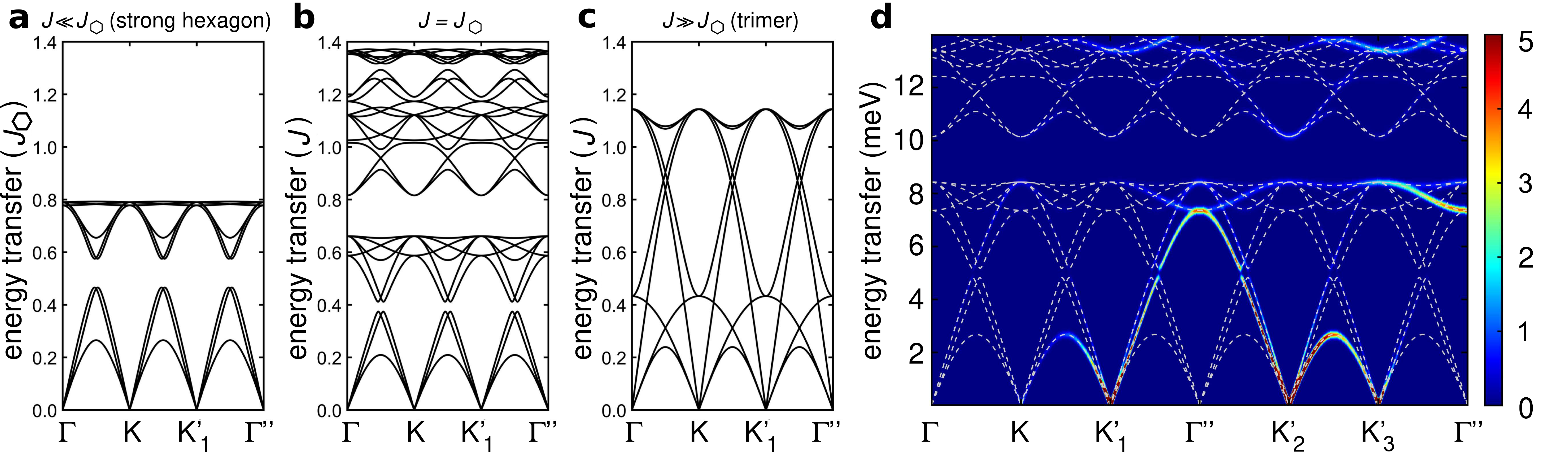}
	\caption{{\bf Spin wave theory results.} Calculated spin-wave dispersion within linear spin wave theory along the $(\xi,\xi)$-direction [from $\Gamma$ to $\Gamma''$, in Fig.~\ref{fig:ninja_lat}\,b] for three different cases: {\bf a} $J/J_{\varhexagon}=1/5$, {\bf b} $J/J_{\varhexagon}=1$, {\bf c} $J/J_{\varhexagon}=25$, where we set $J^\prime = 0$ in each case. For all the choices of couplings considered here, the system is in the $\vec{Q}=(1/3,1/3)$ ordered phase. The energy scale is set by $J_{\varhexagon}$ in panel ({\bf a}), and by $J$ in panel ({\bf b}) and ({\bf c}). We note that in the $J \ll J_{\varhexagon}$ ({\bf a}) and $J = J_{\varhexagon}$ ({\bf b}) cases the lowest band (three-times folded) is separated from the higher bands by an energy gap. This gap closes upon increasing the ratio $J/J_{\varhexagon}$. {\bf d} Calculated spin-wave dispersion and intensity within linear spin wave theory along the path  $\Gamma$-$K$-$K_1^\prime$-$\Gamma^{\prime \prime}$-$K_2^\prime$-$K_3^\prime$-$\Gamma''$ [see Fig.~\ref{fig:ninja_lat}\,b] for the \textit{ab initio} estimated Heisenberg couplings $J = 154.4$~K, $J_{\varhexagon}=134.2$~K  and $J^\prime=8.7$~K ({\SX} structure).%, see Tab.~\ref{tab:2017couplings}).
    The spectral intensity is given by the perpendicular component of the spin dynamical structure factor, which is related to the cross section of unpolarized neutron scattering experiments~\cite{bramwell2011} (the overall color scale is in arbitrary units). The low-energy spectral intensity at the $K^{\prime}_1$ ($K^{\prime}_3$) point is approximately $40\%$ ($20\%$) of the maximal intensity, located at the $K^{\prime}_2$ point. We apply a Gaussian broadening with a standard deviation of $0.2$\,meV. 
	\label{fig:spinwaves_1}}
\end{figure*}

As it represents a previously unexplored magnetic state, it is interesting to study the classical spin wave dispersion of the $\vec{Q}=(1/3,1/3)$ order. In Fig.~\ref{fig:spinwaves_1}, we show the spin wave spectra for the $\vec{Q}=(1/3,1/3)$ magnetic order in three paradigmatic regimes (with $J'=0$ in each case): $J \ll J_{\varhexagon}$ (strong hexagon limit) [Fig.~\ref{fig:spinwaves_1}\,a], $J = J_{\varhexagon}$ [Fig.~\ref{fig:spinwaves_1}\,b],
and $J \gg J_{\varhexagon}$ (strong trimer limit) [Fig.~\ref{fig:spinwaves_1}\,c]. In all three cases the spin wave spectrum has gapless modes at $\Gamma$ and $K$ points. For $J\ll J_{\varhexagon}$, $J = J_{\varhexagon}$, we observe a finite gap between the low-lying magnon band (which gives rise to three branches when folded) and the higher bands. In the strong hexagon limit, where the system is made of weakly coupled hexagons forming a triangular pattern, the excitation spectrum at low energies resembles that of the triangular lattice antiferromagnet (see Supplementary Note 2 and Supplementary Fig.\,3).
The gap between the low-lying branches and the higher ones closes when the ratio $J/J_{\varhexagon}$ is sufficiently large, as shown by the $J\gg J_{\varhexagon}$ case
[Fig.~\ref{fig:spinwaves_1}\,c]. In this limit (strong trimer limit), the system is described by trimers
of spins forming an effective kagome lattice, and the spin wave spectrum resembles that of the kagome $\sqrt{3}\times\sqrt{3}$ magnetic order (see Supplementary Note 2 and Supplementary Fig.\,3).

\begin{figure*}
\centering
\includegraphics[width=0.8\textwidth]{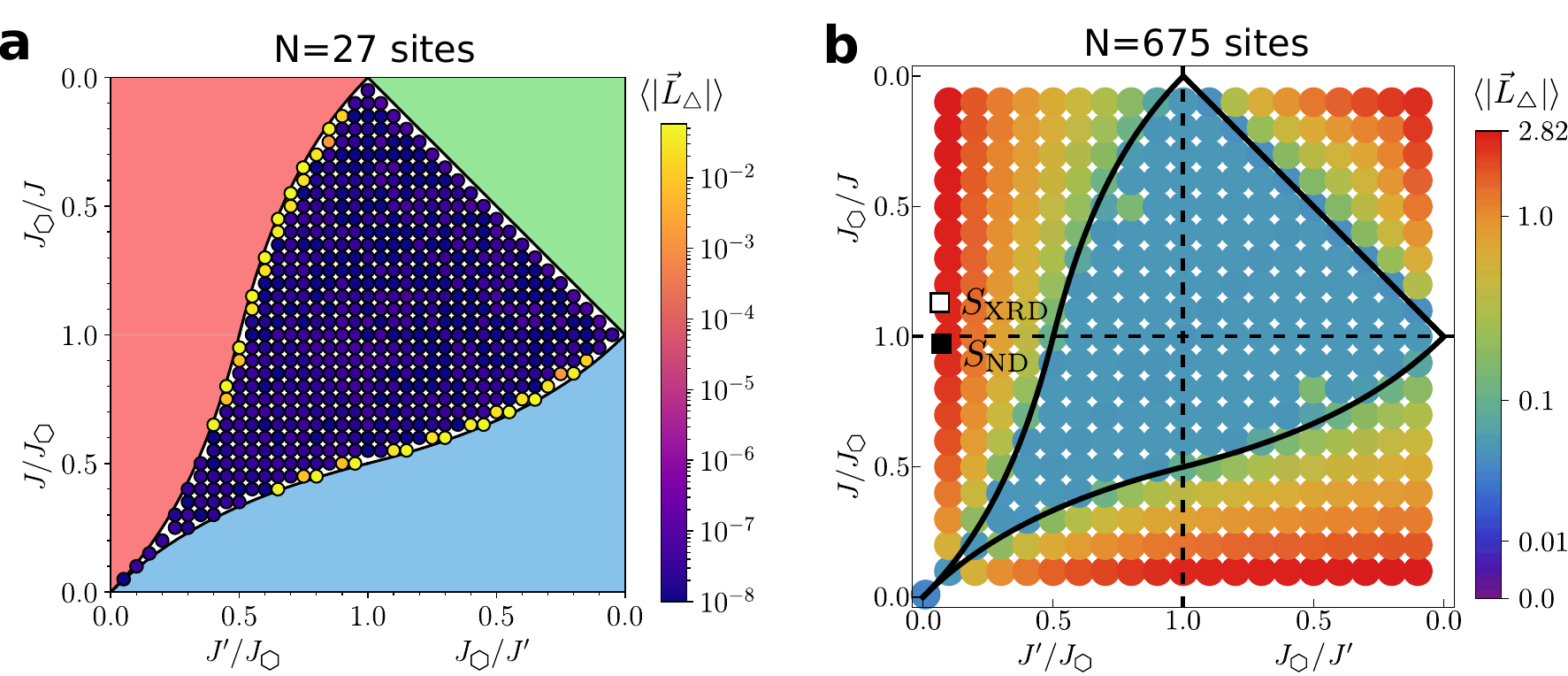}
\caption{{\bf Local ground state constraint within the classical spin liquid phase.} {\bf a} Value of $\langle |\vec{L}_\triangle| \rangle$ (averaged over triangles) in the optimal classical ground state found by iterative minimization. The results are obtained in the classical spin liquid phase on a $N=27$ sites cluster (see Supplementary Fig. 1). Most of the points fulfill the $\vec{L}_{\triangle}=0$ constraint for the ground state. The small number of points with $\langle| \vec{L}_{\triangle}|\rangle>0$ are found to have a coplanar non-degenerate ground state which corresponds to the nearby $\vec{Q}=(1/3,1/3)$ order. 
{\bf b} Phase diagram from classical Monte Carlo simulations. We depict the average of $|\vec{L}_{\triangle}|$ for 10 simulated systems with $N=675$ sites (type-II cluster, $L=5$, cfr. Supplementary Note 1) at the temperature $T=0.001\,J_{\text{max}}$, where ${J_{\text{max}}=\max(J_{\varhexagon},J,J')}$. The $\vec{Q}=(1/3,1/3)$ ($\vec{Q}=0$) ordered phase, with maximal spin susceptibility at the $K'_2$ or $K'_3$ ($\Gamma$) points in the Brillouin zone, lies outside the region defined by Eqs.~\eqref{eq:NonCPCondition} where $\vec{L}_{\triangle}$ can never be zero. Inside the region where $\vec{L}_{\triangle}$ can potentially vanish, we still find finite but small values. The maximum value in the plot, ${\langle |\vec{L}_{\triangle}| \rangle_{\text{max}}\simeq 2.83}$, is found at the points of maximal distortion, i.e. $J'\,(J)=10 \,J_{\varhexagon}=100 \,J\,(J')$. The logarithmic color function scales as $\ln(100\langle |\vec{L}_{\triangle}|\rangle+1)/\ln(100\langle |\vec{L}_{\triangle}|\rangle_{\text{max}}+1)$.
The empty and filled squares indicate two possible sets of couplings for Y-kapellasite. 
\label{fig:iterativeL2_27}}
\end{figure*}

{\bf Classical spin liquid phase.} Finally, we consider the regime where the three couplings generally enable the fulfillment of $\vec{L}_{\triangle}=0$ in Eq.~(\ref{def_l}) and we unveil the ground state nature of this intriguing phase. Even though satisfying $\vec{L}_{\triangle}=0$ in an isolated triangle is possible, this does not immediately imply that achieving $\vec{L}_{\triangle}=0$ in each individual triangle of the full system is a trivial task. In Ref.~\cite{Moessner2017} a generic bond disordered kagome system was investigated for which it was shown that $\vec{L}_{\triangle}=0$ can be satisfied in each triangle. Furthermore, the authors constructed global ground states where each triangle may realize up to two possible spin configurations that locally obey $\vec{L}_{\triangle}=0$ resulting in an extensively, but discretely degenerate classical spin liquid forming a set of the ground states with the cardinality $\aleph_0$, which was named ''jammed spin liquid''.

We performed iterative minimization of the classical energy to confirm the presence of a degenerate manifold of non-coplanar ground states with $\vec{L}_{\triangle}=0$ within the gray region of Fig.~\ref{fig:Classical-PD} (see Methods). The main results of the minimization are summarized in Fig.~\ref{fig:iterativeL2_27}\,a,  where we plot the value of $\langle| \vec{L}_\triangle| \rangle$ (averaged over triangles) as a function of the exchange couplings, for a finite-size cluster with $N=27$ sites (see Supplementary Fig. 1, and Supplementary Fig. 3 for analogous results on a $N=36$ site cluster). In the optimal solutions, $|\vec{L}_\triangle|$ is actually found to be identical for all triangles. Its square value yields the residual energy per triangle with respect to the ideal ground state with $\vec{L}_\triangle=0$. We observe that in most of the region delimited by the boundaries of Eq.~\eqref{eq:NonCPCondition}, we obtain a \emph{degenerate} set of ground states with $|\vec{L}_\triangle|^2=0$ for each triangle (within machine precision). Indeed, starting the iterative minimization with different initial points, we find several independent minima which cannot be connected to each other by lattice symmetries and global spin rotations, thus confirming the large degeneracy of the classical ground state, as predicted in Ref.~\cite{Moessner2017}. We also note that the $\vec{L}_\triangle=0$ solutions found by the minimization are, in general, non-coplanar, and can be exploited to construct $\vec{L}_\triangle=0$ ground states for larger systems, by simply using the $N=27$ sites cluster as effective unit cells for type-II clusters (see Supplementary Note 1).% (cfr. Appendix~\ref{sec:app-finite}).
    
However, close to the boundaries of the non-coplanar region, we find a number of points in the phase diagram where achieving $\vec{L}_\triangle =0$ by numerical minimization was not possible. We are not able to provide a final statement whether the $\langle |\vec{L}_{\triangle}|\rangle>0$ points close to the boundaries are an artifact of the finite-size calculations (and boundary conditions~\cite{Moessner2017}), or whether they belong to the neighboring $\vec{Q}=(1/3,1/3)$ ordered region, which may extend slightly beyond the ideal boundaries of Eq.~\eqref{eq:NonCPCondition}.
Indeed, it is worth noting that the best $\vec{Q}=(1/3,1/3)$ solution at the analytical phase boundary with the spin-liquid phase [i.e., when the equality of Eq.~(\ref{eq:NonCPCondition}a) or~(\ref{eq:NonCPCondition}b) holds] always has a finite residual energy, i.e. $\langle|\vec{L}_{\triangle}|\rangle>0$ (except for the extreme cases where one coupling is zero). Thus, a continuous deformation of the $\vec{Q}=(1/3,1/3)$ order to the $|\vec{L}_{\triangle}|=0$ spin-liquid phase cannot take place at the precise location of the analytical boundaries. Therefore, either the transition is not continuous, or the position of the phase boundaries is slightly shifted with respect to the analytical conditions of Eqs.~(\ref{eq:NonCPCondition}a) and~(\ref{eq:NonCPCondition}b).
Nevertheless, except for the precise location of the transition, our numerical minimization confirms the presence of an extended classical spin-liquid phase, characterized by degenerate non-coplanar ground states.

\begin{figure*}
	\centering
	\includegraphics[width=0.8\textwidth]{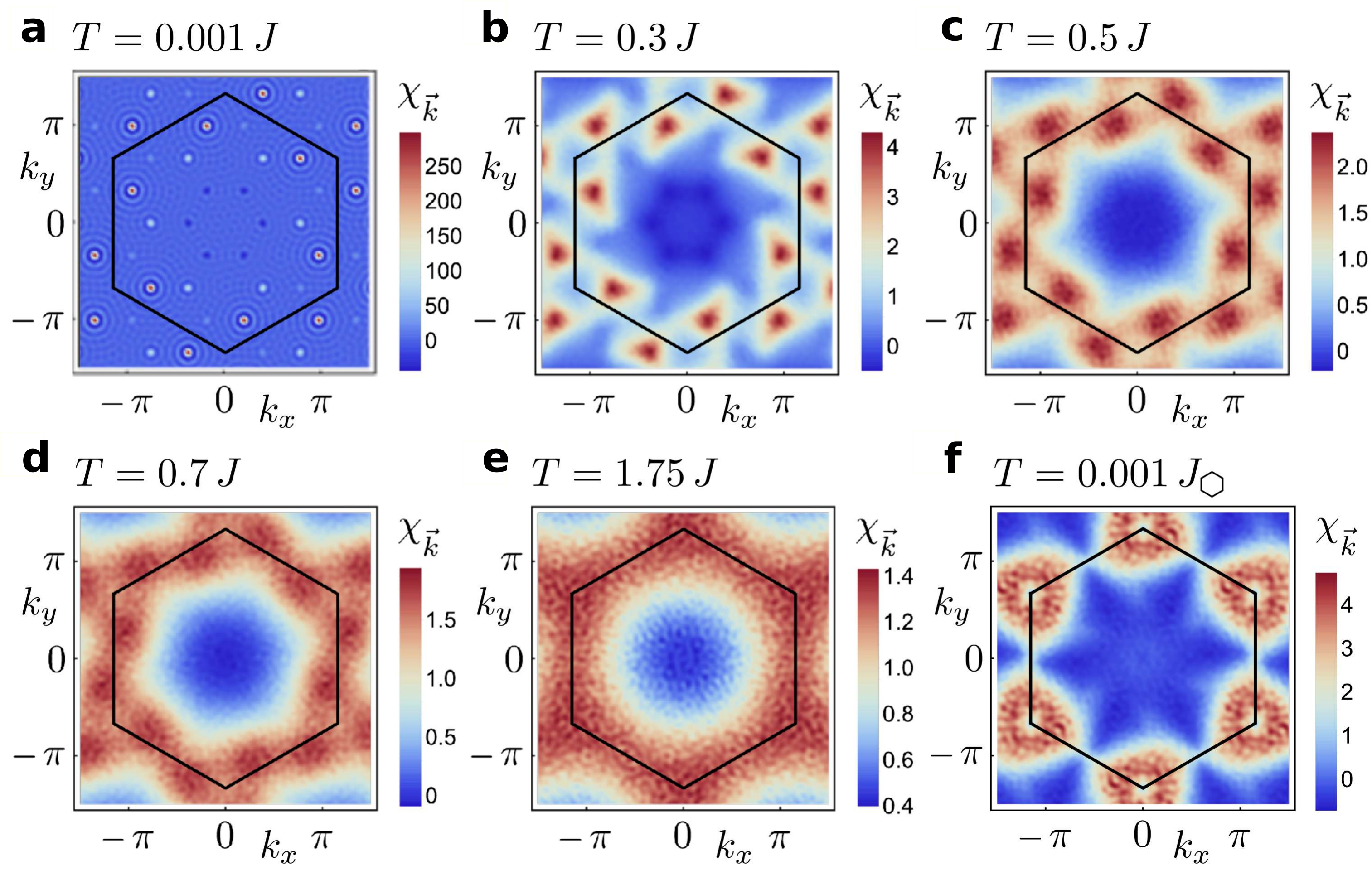}
	\caption{{\bf Classical Monte Carlo results.} {\bf a} to {\bf e}: Spin susceptibility in momentum space at different temperatures from classical Monte Carlo simulations for the {\SX} structure on a cluster with $N=7803$ spins (type-II cluster with $L=17$, see Supplementary Note 1), with $\gamma=0.001$ (see Methods %Appendix~\ref{sec:CMCappendix}
    for details). We only consider finite real-space correlations within a circle with a radius of $50$ nearest-neighbor distances around each spin. The extended Brillouin zone is depicted as a black hexagon, cfr. Fig.~\ref{fig:ninja_lat}\,b. {\bf f} Spin susceptibility for $J/J_{\varhexagon}=0.5$ and $J'/J_{\varhexagon}=0.45$, within the non-coplanar phase. The susceptibility has been computed by classical Monte Carlo calculations for ten ${N=7803}$ sites clusters (type-II, $L=17$, see Supplementary Note 1) at $T=0.001\,J_{\varhexagon}$. \label{fig:CMC-susc}}
\end{figure*}

In addition to energy minimization, we performed classical Monte Carlo simulations in the low temperature limit (see Methods),%Appendix~\ref{sec:CMCappendix} for the calculations details),
computing the value of $\langle |\vec{L}_{\triangle}|\rangle$
in the full phase diagram, as shown in Fig.~\ref{fig:iterativeL2_27}\,b. The Monte Carlo results confirm the presence of a region of non-coplanar ground states within the boundaries of Eq.~\eqref{eq:NonCPCondition}. In this region the value of $\langle|\vec{L}_{\triangle}|\rangle$ is found to be clearly smaller than in the rest of the phase diagram, where the $\vec{Q}=(1/3,1/3)$ and $\vec{Q}=0$ orders are observed. The finite value of $\langle|\vec{L}_{\triangle}|\rangle$ within the non-coplanar region can be ascribed to the effect of finite temperature. To further characterize the properties of this phase, we compute the spin susceptibility [Eq.~\eqref{eq:SpinSusceptibility}]. As shown in Fig.~\ref{fig:CMC-susc}~{f}, the spin-spin correlations in the non-coplanar phase cannot be described by any particular wave vector, but rather by a distribution of wave vectors.

\begin{figure*}
	\includegraphics[width=0.9\textwidth]{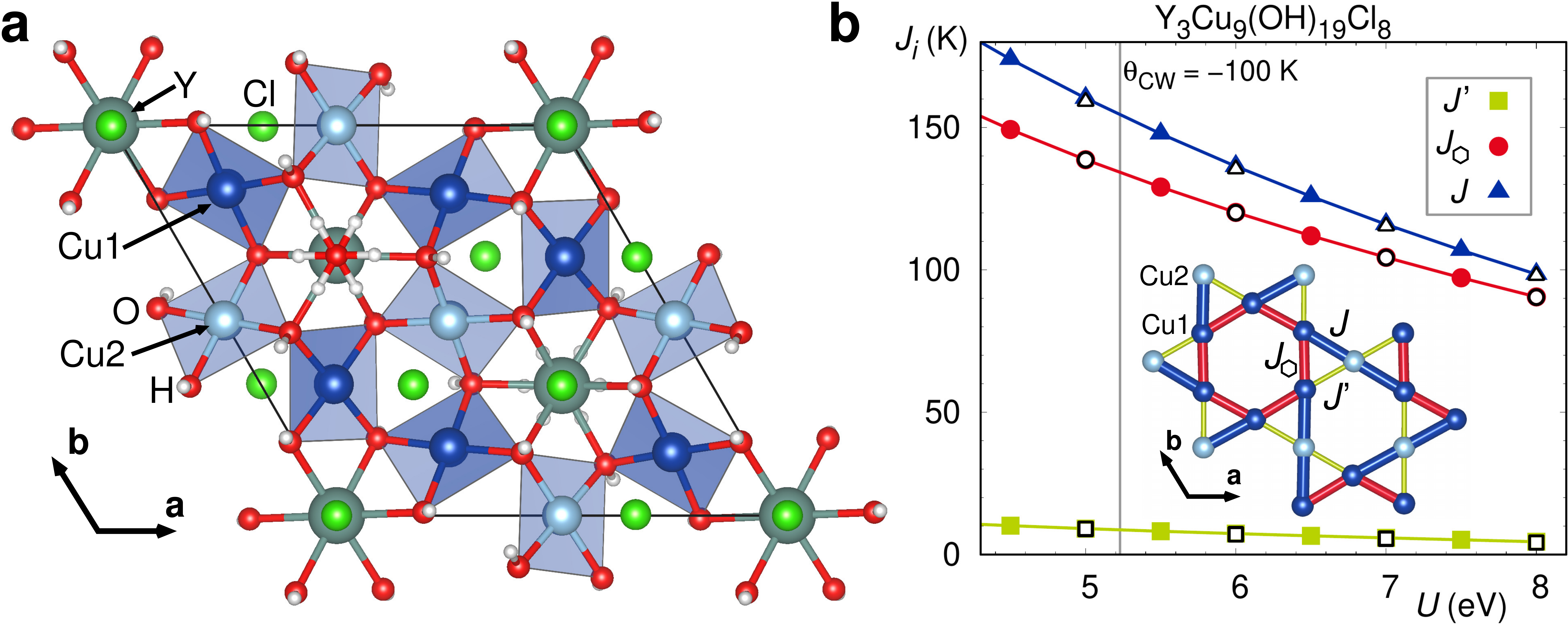}
	\caption{{\bf Structure and exchange couplings of {\y}.} {\bf a} Crystal structure of Y-kapellasite~\cite{Puphal2017} (space group 148, $R\bar{3}$) with DFT relaxed hydrogen positions. {\bf b}  Exchange couplings of {\y} as function of interaction strength $U$, determined by energy mapping using GGA+U at $J_H=1$~eV . Solid symbols: $P1$ cell, 7 couplings extracted. Empty symbols: $\sqrt{2}\times\sqrt{2}\times 1$ supercell, 24 couplings extracted. The inset shows the nearest-neighbor exchange paths of the perfect kagome lattice which differentiate into $J_{\varhexagon}$, $J$ and $J'$ in Y-kapellasite.\label{fig:couplings}}
\end{figure*}

{\bf Magnetic Hamiltonian of Y-kapellasite.}
We now concentrate on the specific case of Y-kapellasite and investigate the magnetic properties of its spin Hamiltonian in more detail, also including the effects of quantum fluctuations. We start by performing \textit{ab initio} density functional theory calculations to confirm that Y-kapellasite supports the spin model of Fig.~\ref{fig:ninja_lat}\,a and determine the precise values of the coupling constants.
We used both published crystal structures, the one determined by X-ray diffraction of single crystals~\cite{Puphal2017},
and the structure of {\yd} determined by neutron diffraction on powder samples~\cite{Barthelemy2019}. We consider the former structure more reliable than the latter (we follow the privately communicated assessment of P. Puphal that the single crystal samples of Ref.~\cite{Puphal2017} are more pure and less strained than the deuterated powders of Ref.~\cite{Barthelemy2019}), and we will refer to the single crystal structure~\cite{Puphal2017} as {\SX} structure and to the powder structure~\cite{Barthelemy2019} as {\SN} structure (see Methods for more details). Note that our analysis below is valid for both structures.

The three largest couplings $J$, $J_{\varhexagon}$ and $J^{\prime}$ (all antiferromagnetic) are shown in Fig.~\ref{fig:couplings}\,b, as obtained by energy mapping for the {\SX} structure of Y-kapellasite. The couplings are tabulated in Supplementary Table 1. %Appendix~\ref{sec:dftappendix}, Table~\ref{tab:2017couplings}.
These three couplings form the distorted kagome lattice illustrated in the inset of Fig.~\ref{fig:couplings}\,b. 
Our reasoning that the relevant exchange couplings for Y-kapellasite are just the three nearest neighbours of the distorted kagome lattice is based on extensive energy mapping for seven and, for additional confidence, 24 neighbours up to Cr-Cr distances of 8.14\,{\AA}.
The determination of 24 couplings for a larger supercell as listed in Supplementary Table 3 % Table~\ref{tab:2019allcouplings} of Appendix~\ref{sec:dftappendix},
fully confirms the three largest couplings, shown as empty symbols in Fig.~\ref{fig:couplings}\,b. It also shows that Y-kapellasite is a very two-dimensional material, and in the following we neglect all interlayer couplings. Among the three second and six third nearest neighbor couplings of the distorted kagome lattice, the largest is $J_7$ with a value 2{\%} of $J$, which is rather small. This means that it is justified to focus the study of Y-kapellasite on the nearest-neighbor Hamiltonian.
The couplings for the {\SN} structure of Y-kapellasite are given in Supplementary Fig.\,8 and Supplementary Table 2, respectively. % Fig.~\ref{fig:2019couplings} and Table~\ref{tab:2019couplings} in Appendix~\ref{sec:dftappendix}, respectively.
There is one clear difference between {\SX} and {\SN}: $J_{\varhexagon}$ is 13{\%} smaller than $J$ for the {\SX} structure but 3{\%} larger for the {\SN} structure. We will discuss the implications for the Hamiltonian in the next section.
We emphasize that, according to the \textit{ab initio} calculations above, the value of the exchange coupling $J'$ is considerably smaller than those of $J_{\varhexagon}$ and $J$, which are of comparable size. 
In conclusion, by calculating and inspecting a large number of exchange couplings, we verified that the $J$, $J_{\varhexagon}$, $J^{\prime}$ Hamiltonian is not a simplifying assumption but a defining feature of the material Y-kapellasite.

In what follows, we concentrate on the magnetic properties of Y-kapellasite as described by the structure {\SX} and the coupling constants $J$, $J_{\varhexagon}$, $J'$ of Supplementary Table 1 %Table~\ref{tab:2017couplings}
which place the material in the $\vec{Q}=(1/3,1/3)$ ordered regime of the classical phase diagram (see Fig.~\ref{fig:Classical-PD}). This placement puts our calculations in agreement with very recent experimental observation of (partial) magnetic order in Y-kapellasite~\cite{Sun2021}.

{\bf Classical Monte Carlo simulations.}
We start our investigation of the Heisenberg Hamiltonian of Y-kapellasite (for the {\SX} structure) using the classical Monte-Carlo technique. Despite neglecting quantum fluctuations, this analysis allows us to study how thermal fluctuations impact the $\vec{Q}=(1/3,1/3)$ order (see Methods for technical details). %. Technical details of the calculations are given in Appendix~\ref{sec:CMCappendix}.
In Fig.~\ref{fig:CMC-susc} we present results on the spin susceptibility in momentum space [see Eq.~(\ref{eq:SpinSusceptibility})] for different temperatures. At high temperatures [Fig.~\ref{fig:CMC-susc}~{e}], the response is almost homogeneous along the edges of the extended Brillouin zone. This response resembles the one of the standard undistorted nearest-neighbor kagome model, indicating that at these high temperatures details of the precise detuning between $J$, $J_{\varhexagon}$ and $J'$ do not yet affect the susceptibility. When $T$ is lowered [going from panel {e} to panel {a} in Fig.~\ref{fig:CMC-susc}], additional features become discernible such as three maxima around each corner of the extended Brillouin zone ($\sqrt{3}\times\sqrt{3}$ positions). Each such triad forms an equilateral triangle and with decreasing temperature the peaks become sharper. Simultaneously, the triangles show a slight rotation around their center points ($\sqrt{3}\times\sqrt{3}$ positions) until in the low-temperature limit the peaks reach the $\vec{Q}=(1/3,1/3)$ order positions [$K'_2$ points in Fig.~\ref{fig:ninja_lat}\,b]. This shift of peaks roughly occurs along a line connecting the $K'_2$ and $\Gamma''$ points. Please see Fig.~\ref{fig:FRGFlows}\,c for a trace of the peak positions at different temperatures. Note that as a result of the Mermin-Wagner theorem, real long-range magnetic order is possible only at strictly $T=0$. However, the fact that the short-range correlations in the intermediate temperature regime manifest themselves in susceptibility peaks at incommensurate wave vectors away from $\vec{Q}=(1/3,1/3)$ indicates that thermal fluctuation act in a non-trivial and unexpected way. At T=0, where the $\vec{Q}=(1/3,1/3)$ sets in, the maxima of the susceptibility are located at the $K^{\prime}_2$ points, with smaller peaks appearing at the $K_1'$ and $K_3'$ points. The relative heights of the latter peaks are $\chi_{K_1'}/\chi_{K_2'} \approx 40\%$ and $\chi_{K_3'}/\chi_{K_2'} \approx 20\%$, respectively.

\begin{figure}
	\centering
	\includegraphics[width=0.85\columnwidth]{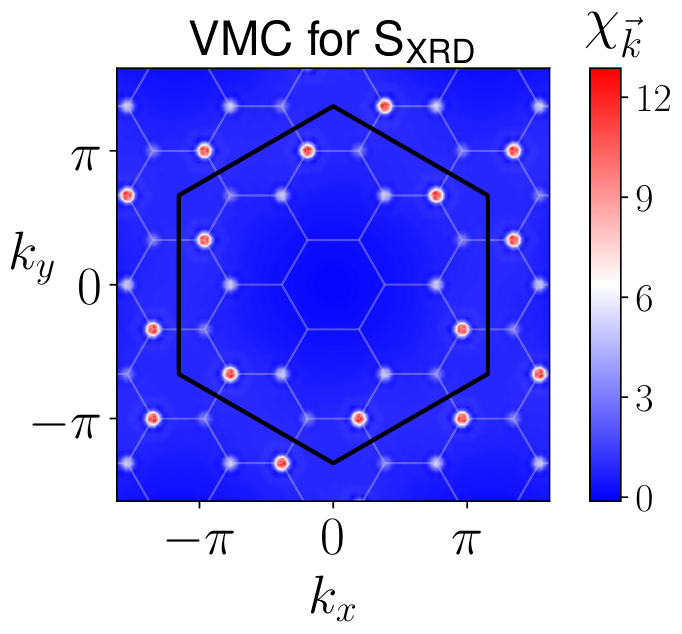}
	\caption{{\bf Variational Monte Carlo result.} Spin susceptibility in momentum space [Eq.~\eqref{eq:SpinSusceptibility}] computed by VMC. The results refer to the optimal variational state for the Hamiltonian of Eq.~\eqref{eq:heis_ham} with the exchange couplings of the {\SX} structure of Y-kapellasite (see Supplementary Table 1). %(see Tab.~\ref{tab:2017couplings}).
    The calculation is performed on a finite cluster of $N=972$ sites (type-II, $L=6$, cfr. Supplementary Note 1).
	\label{fig:vmc_sq}}
\end{figure}

\begin{figure*}
	\centering
	\includegraphics[width=0.95\textwidth]{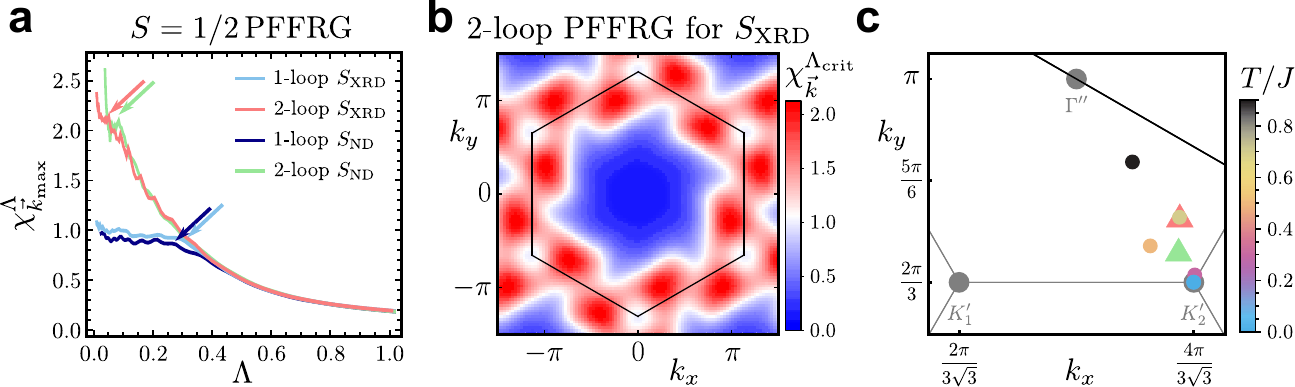}
	\caption{{\bf Pseudofermion functional renormalization group results.} {\bf a} One- and two-loop spin susceptibility flows from PFFRG for {\SX} and {\SN} structures. The arrows mark a kink or peak during the flow indicating the onset of magnetic order. {\bf b} Spin susceptibility in $\vec{k}$ space [Eq.~\eqref{eq:SpinSusceptibility}] from two-loop $S=1/2$ PFFRG for {\SX} structure at the critical cutoff (marked by red arrow in {\bf a}). The extended Brillouin zone is indicated by the black hexagon. Maxima of $\chi_{\vec{k}}^{\Lambda_{\text{crit}}}$ appear at incommensurate positions. {\bf c} Momentum-space position of the maximal susceptibility at the breakdown of the PFFRG flow together with the corresponding values obtained from classical Monte Carlo at finite temperature $T$. We show a section of the first Brillouin zone (gray lines) close to one edge of the extended Brillouin zone (black line), confer Fig.~\ref{fig:ninja_lat}\,b. The PFFRG results for the {\SX} parameters are shown as triangles (the peak positions for the {\SN} structure are similar). Red and green triangles represent one-loop and two-loop results, respectively. Dots represent the susceptibility maxima from classical Monte Carlo for the system with {\SX} couplings and the color represents temperature.\label{fig:FRGFlows}}
\end{figure*}

{\bf Variational Monte Carlo results.}
We now analyze the ground state properties of the spin model in the quantum regime with variational Monte Carlo (VMC) calculations. As in the previous section, we focus on the set of exchange couplings obtained for the {\SX} structure of Y-kapellasite, which lies in the $\vec{Q}=(1/3,1/3)$ ordered region of the classical phase diagram (see Fig.~\ref{fig:Classical-PD}). Our variational method is based on Gutzwiller-projected fermionic states (see Methods and Supplementary Note 5 for details). %, is described in detail in Appendix~\ref{sec:vmc-app}.
Optimizing the variational state, we obtain a finite value of the magnetic field variational parameter, which indicates the resilience of the classical $\vec{Q}=(1/3,1/3)$ magnetic order against quantum fluctuations. To corroborate this finding, we compute the spin susceptibility, Eq.~\eqref{eq:SpinSusceptibility}, with ${\langle \cdots \rangle=\langle \Psi_0 | \cdots | \Psi_0 \rangle}$ representing the expectation value over the optimal variational wave function. The results for a finite cluster of $N=972$ sites (type-II, $L=6$, cfr. Supplementary Note 1) %Appendix~\ref{sec:app-finite})
are shown in Fig.~\ref{fig:vmc_sq}: the susceptibility is clearly dominated by sharp Bragg peaks at the $K_2'$ points of the extended Brillouin zone, thus confirming the presence of $\vec{Q}=(1/3,1/3)$ magnetic order. We note that $\chi_{\vec{k}}$ is not significantly different from the classical result at zero temperature ($i.e.$, no specific features are detected except for the Bragg peaks), despite the important contributions of the fermionic hoppings and the Jastrow factor to the variational energy. An almost identical susceptibility is obtained when considering the exchange couplings of the {\SN} structure. Thus, according to our VMC results, the minimal Heisenberg model for Y-kapellasite has a $\vec{Q}=(1/3,1/3)$ magnetically ordered ground state.

{\bf Pseudofermion functional renormalization group calculations.}
Next, we employ the pseudofermion functional renormalization group (PFFRG) approach~\cite{Reuther2010,Rueck2018,Baez2017, Buessen18,Buessen19} to investigate ground state quantum effects in our distorted kagome Heisenberg model from a complementary methodological perspective. Within PFFRG, we compute the static spin susceptibility in momentum space $\chi^{\Lambda}_{\vec{k}}$ as a function of the RG parameter $\Lambda$ (which acts as a low-energy frequency cutoff). We employ two variants of this technique, the one-loop and two-loop schemes, where the latter can be considered more accurate (but computationally more demanding) as it includes additional diagrammatic contributions to better account for the system's fluctuations beyond mean-field (see Methods for details). Most importantly, an onset of magnetic order can be observed as an instability during the RG flow of the maximal $\vec{k}$-space component of $\chi^{\Lambda}_{\vec{k}}$. Such an instability is indeed evident in the RG flows for both schemes (one-loop, two-loop) and for both structures of Y-kapellasite (see Fig.~\ref{fig:FRGFlows}\,a) confirming the findings from VMC. However, the fact that these instability features are quite weak and only detectable as small kinks rather than sharp peaks indicates the significance of quantum fluctuations, possibly associated with a small ordered moment. Note that the instability is observed at much smaller $\Lambda$ in the two-loop scheme as compared to the one-loop approach, which is a known property resulting from the better fulfillment of the Mermin-Wagner theorem in the former method~\cite{Rueck2018}.
The momentum resolved susceptibility $\chi^{\Lambda_\text{crit}}_{\vec{k}}$ at the critical RG scale $\Lambda_\text{crit}$ from two-loop PFFRG for the {\SX} structure is shown in Fig.~\ref{fig:FRGFlows}\,b. The maxima are rather broad, again indicating strong effects of quantum fluctuations. Furthermore, the peaks do not exactly coincide with the $\vec{Q}=(1/3,1/3)$ positions, as in VMC results, but show a small shift along the $K'_{2}-\Gamma''$-line, resembling our above findings from classical Monte Carlo. This indicates that quantum fluctuations may have similar effects as thermal fluctuations. We emphasize that this behavior is rather unusual, since, typically, quantum fluctuations tend to lock magnetic orders at commensurate positions. 
In Fig.~\ref{fig:FRGFlows}\,c, we summarize the peak positions from one-loop and two-loop PFFRG as well as from classical Monte Carlo at intermediate temperatures. As can be seen, all results show a shift along similar momentum-space directions, however, the displacement away from the $\vec{Q}=(1/3,1/3)$ point becomes smaller as we advance the approach from one-loop to two-loop. It is hence conceivable that the shift would completely disappear upon further improving the method towards multi-loop schemes~\cite{thoenniss20,kiese20}. We leave this as an open question for future investigations. We remark, however, that VMC and PFFRG both find magnetic long range order
for Y-kapellasite.

{\bf Linear Spin Wave Theory.}
We conclude the analysis of the magnetic properties of Y-kapellasite by showing in Fig.~\ref{fig:spinwaves_1}\,d the spin wave  spectrum and intensities  for the Heisenberg Hamiltonian corresponding to the {\SX} structure. We observe that the spectrum is very similar to the simpler case of  $J = J_{\varhexagon}$ and $J'=0$ (Fig.~\ref{fig:spinwaves_1}), which can then be regarded as a reliable minimal approximation for the full model. The intensity is largest at the $K^{\prime}_2$ point [see Fig.~\ref{fig:ninja_lat}\,b] corresponding to the $\vec{Q}=(1/3,1/3)$ order, as also observed in the spin susceptibility results above.

\vspace{0.2cm}{\bf\large Discussion}

Summarizing, by a combination of DFT, effective spin models, classical (iterative minimization, classical Monte Carlo) and quantum approaches (VMC, PFFRG) we investigated the magnetic properties of a  distorted kagome lattice as realized in the recently synthesized Y-kapellasite. We found an unexpectedly rich phase diagram already at the classical level  which includes a  collinear $\vec{Q} = 0$ magnetic phase, two unusual non-collinear coplanar $\vec{Q} = (1/3,1/3)$ magnetic phases, and a classical spin liquid phase that resembles the jammed spin liquid phase found in the context of a bond-disordered kagome antiferromagnet. Our analysis of the spin model for Y-kapellasite places this system in the region of $\vec{Q}=(1/3,1/3)$ magnetic order with an excitation spectrum  that lies halfway between  that of an underlying triangular lattice  of hexagons  and a kagome lattice of trimers.

While it is not experimentally settled whether Y-kapellasite orders magnetically, our theoretical results provide strong evidence in favor of a magnetic $\vec{Q}=(1/3,1/3)$ ground state. The presence of an extended classical spin liquid phase in the vicinity of our DFT couplings sheds additional interesting light on this compound. Possibly, through external perturbations such as pressure or strain one might be able to shift the couplings towards the classical spin liquid phase, which, due to the large extent of this regime, may not require any fine-tuning. This opens the question about the fate of the classical spin liquid upon adding quantum fluctuations, which we did not tackle in this work. Given the complexity of this phase already on the classical level one may expect even richer phenomena in the quantum case, including a quantum spin liquid. The numerical investigation of this regime in the quantum limit will certainly be a challenging future task but also gives hope for rewarding insights.
In total, this work demonstrates that a relatively simple but realistic distortion of the kagome lattice gives rise to a multitude of interesting and unexpected magnetic phenomena whose full investigation goes far beyond the scope of the present work. In the future, our investigation may inspire and guide both a deeper experimentally motivated investigation of Y-kapellasite, as well as a closer numerical analysis of the underlying spin model.

\vspace{0.2cm}{\bf\large Methods}

{\bf Density functional theory based energy mapping.}
We calculate the electronic structure of Y-kapellasite 
with DFT using the full potential local orbital (FPLO) basis set~\cite{Koepernik1999} and the generalized gradient approximation (GGA) to the exchange correlation functional~\cite{Perdew1996}. We apply the GGA+U approximation~\cite{Liechtenstein1995} to correct for  strong electronic correlations of the Cu $3d$ electrons. We set the Hund's rule coupling to a typical value value~\cite{Jeschke2013,Jeschke2015} $J_H=1$~eV for Cu$^{2+}$ and vary only the onsite interaction $U$. Even though the  
{\SN} structure~\cite{Barthelemy2019} nominally has 8/9 filling, there is no evidence that Y-kapellasite is charge doped, and therefore we treat the O1 position as occupied with a hydroxy group (or a chloride ion which leads to the same results). In this position, the {\SX} structure~\cite{Puphal2017} has an orientationally disordered OH$^-$ ion between two Y$^{3+}$ ions and therefore a 1/6 occupation of the six symmetry equivalent H positions is consistent with the $R\bar{3}$ space group.
We model the orientationally disordered OH$^-$ ion using the virtual crystal approximation~\cite{nordheim1931}, setting the nuclear charge of H in this position to $1/6$. The hydrogen positions H2 to H4 are relaxed within GGA in both structures. We shift the partially occupied H1 hydrogen position to the equilibrium O-H distance. The resulting structure is shown in Fig.~\ref{fig:couplings}\,a.

We use total energy mapping~\cite{Guterding2016,Iqbal2017}
to determine the Heisenberg Hamiltonian
parameters of Y-kapellasite. For that we calculate with DFT(GGA+U) the total energy for 24 out of the 47 unique spin configurations which are possible with the 9 inequivalent Cu$^{2+}$ ions in the $P\,1$ unit cell of {\y}. Considering that third-neighbor couplings are important for some kapellasite type compounds~\cite{Jeschke2013}, we also perform calculations for a $\sqrt{2}\times\sqrt{2}\times 1$ supercell with 18 independent Cu sites; we calculate 44 out of nearly 30000 spin configurations with distinct energies.

{\bf Iterative minimization.}
To numerically determine the classical ground state of the spin model of Eq.~\eqref{eq:heis_ham}, we employ the iterative minimization method~\cite{walker1980,sklan2013}. We initialize our system in a random configuration and we iteratively perform local moves to update the spins. In each move, we pick up a random spin, $\vec{S}_i$, and we orient it antiparallel to the local field created by the neighboring spins, i.e.
\begin{equation}
\vec{S}_i \mapsto -\frac{\vec{B}_i}{|\vec{B}_i|}, \quad \mbox{with } \vec{B}_i=\sum_{j} J_{ij}\vec{S}_j.
\end{equation}
The procedure is repeated for a sufficient number of steps until the energy converges. In order to reduce the risk of ending up in local minima, we repeat the calculations several times starting from different spin configurations and we keep the solution with the best energy. The calculations are performed on the small finite-size clusters shown in Supplementary Fig. 1 %Fig.~\ref{fig:iterativeclu},
with $N=27$ and $N=36$ sites, and periodic boundary conditions.
It is important to emphasize that finding a classical ground state with $\vec{L}_\triangle=0$ on one of these small clusters (with periodic boundary conditions) implies that one can immediately define a $\vec{L}_\triangle=0$ ground state for any larger cluster of the same type (see Supplementary Note 1). %(see Appendix~\ref{sec:app-finite}). 

{\bf Classical Monte Carlo simulations.}
We perform a Monte Carlo analysis using the Metropolis algorithm with over-relaxation protocol for better thermal convergence~\cite{creutz87,kanki05,zhitomirksy08,pixley08}. 
 For the investigations of the {\SX} structure,
 the system that we simulate is a cluster of $N=7803$ spins with periodic boundary conditions (type-II cluster with $L=17$ (Supplementary Fig. 1). %cfr. Appendix~\ref{sec:app-finite}).
 It is seeded at $T_0=2 \, J_{\text{max}}$ with random spins and cooled down via $T=T_0e^{-\gamma n}$ where $n=0,\,1,\,2,\,\dots$ is the number of Metropolis steps.
During each step, every spin is updated once on average by a new random spin. The update takes place either with certainty if the acquired energy $\Delta E \leq 0$ or with a probability $p=e^{-\Delta E/T}$. For the different coupling regimes, we set $\gamma=0.001$ and cooled $100$ ($10$) random systems down to $T=0.1 \, J_{\text{max}}$ ($T=0.001 \, J_{\text{max}}$), where ${J_{\text{max}}=\max(J_{\varhexagon},J,J')}$. For the investigation of the classical phase diagram we 
  restrict ourselves to type-II clusters with $L=5$ ($N=675$ spins), for a numerical speedup.

{\bf Linear spin wave theory.} We performed linear spin wave calculations with the \textsc{SpinW} package~\cite{spinwref}, computing the classical ground state by energy minimization.

{\bf Variational Monte Carlo.}
We employ Gutzwiller-projected fermionic states as variational \textit{ans\"atze}~\cite{Becca2017}. This class of wave functions has been shown to provide an accurate description of the ground state of several spin models~\cite{Becca2011}, including state-of-the-art results for kagome lattice antiferromagnets~\cite{Iqbal2013,Iqbal2015,Iqbal2018,Zhang2020,Ferrari2021}. The optimal variational \textit{ansatz} for the Hamiltonian of Y-kapellasite is a Gutzwiller-projected Jastrow-Slater wave function possessing $\vec{Q}=(1/3,1/3)$ magnetic order (see Supplementary Note 5 for the definition). For the optimization of the variational parameters we use the stochastic reconfiguration method~\cite{sorella1998}.

{\bf Pseudofermion Functional Renormalization Group.}
The PFFRG method is based on the one-loop plus Katanin truncation PFFRG scheme first introduced in Ref.~\onlinecite{Reuther2010} and extended to the two-loop plus Katanin variant in Ref.~\onlinecite{Rueck2018}. It utilizes the Abrikosov pseudofermion representation of spin operators. This spin representation enlarges the Hilbert space by adding two unphysical $S=0$ states per site (unoccupied, doubly occupied) which, however, leave the ground state properties largely unaffected~\cite{Baez2017}. Within PFFRG the bare propagator of the fermions is regularized by a sharp cutoff function:
\begin{equation}\label{eq:GCutoff}
    G_0(\omega)=\frac{1}{i\omega} \quad \longrightarrow \quad G_0^{\Lambda}(\omega)=\frac{\theta(|\omega|-\Lambda)}{i\omega}\;.
\end{equation}
Here, $\omega$ is a continuous Matsubara frequency at $T=0$ and the cutoff $\Lambda$ prohibits fermionic propagation if $|\omega|\leq\Lambda$. This insertion causes a cutoff dependence of the generating functional for the fermionic one-particle-irreducible vertex functions. Flow equations which describe the $\Lambda$ derivatives of all $n$-particle vertex functions can be derived. These equations couple the $n$-particle vertex to all $m$-particle vertices with $m\leq n+1$ leading to an infinite hierarchy of equations. In principle, physical results in the cutoff-free limit $\Lambda\rightarrow 0$ can be obtained by solving the integro-differential flow equations starting from the limit $\Lambda\rightarrow\infty$ where the initial conditions are set by the bare interactions from our spin Hamiltonian. For numerical solvability this hierarchy of equations needs to be truncated. In the one-loop scheme, the truncation occurs on the level of the three-particle vertex which is replaced by contributions from the Katanin scheme~\cite{Katanin2004}, particularly, the single-scale propagator is upgraded to
$S^{\Lambda}(\omega)=-\frac{d}{d\Lambda}G^{\Lambda}(\omega)$ where the full Green's function is $G^{\Lambda}(\omega)=\left[\left(G_{0}^{\Lambda}(\omega)\right)^{-1} - \Sigma^{\Lambda}(\omega) \right]^{-1}$. The one-loop flow equations for the self energy $\Sigma^{\Lambda}$ and the two-particle vertex $\Gamma^{\Lambda}$ are depicted diagrammatically in Supplementary Fig.\,7. %Fig.~\ref{fig:FlowEquations}.
In the two-loop approach further contributions of the three-particle vertex are included, which have the form of nested one-loop diagrams~\cite{Rueck2018}. We solve the flow equations numerically with an Euler scheme in real space taking into account finite spin correlations on hexagonal clusters with an edge length of $N\geq 7$ nearest-neighbor distances around reference sites from each sublattice. The Matsubara frequencies are discretized using a linear plus logarithmic mesh with $M_{\omega} \geq 60$ points. We carefully analyzed that the qualitative PFFRG results are converged with respect to the number of frequency points and the finite correlation length. From the resulting two-particle vertex, we are able to compute the $\Lambda$ dependent static spin susceptibility $\chi^{\Lambda}_{\vec{k}}$ in momentum space. For more details we refer the reader to Refs.~\onlinecite{Reuther2010,Baez2017,Rueck2018}.

\vspace{0.2cm}
{\bf DATA AVAILABILITY}\\
The datasets generated and/or analysed during the current study are available from the corresponding authors upon reasonable request.

\vspace{0.2cm}
{\bf CODE AVAILABILITY}\\
The calculation codes used in this paper are available from the corresponding authors upon reasonable request.

\vspace{0.2cm}
{\bf\large Acknowledgments}\\
We thank Q.~Barthélemy, F.~Bert, T.~Biesner, M.~Dressel, E.~Kermarrec, P.~Mendels, P.~Puphal, S.~Roh for useful discussions. F.F. acknowledges support from the Alexander von Humboldt Foundation through a postdoctoral Humboldt fellowship. A.R., R.V. and J.R. acknowledge support by the Deutsche Forschungsgemeinschaft (DFG, German Research Foundation) for funding through TRR 288 - 422213477 (projects A05, B05) (A.R. and R.V.) and CRC 183 (project A04) (J.R.).  I.I.M. acknowledges  support from the U.S. Department of Energy through Grant No. DE-SC0021089 and from the Wilhelm and Else Heraeus Foundation.
\vspace{0.2cm}

{\bf\large Author contributions}\\ M.H. and F.F. contributed equally to this work. R.V. and H.O.J. conceived the project. I.I.M., R.V., H.O.J. and J.R. supervised the project. The analytical calculations were performed by F.F., A.R. and I.I.M. The iterative minimization was done by F.F., the classical Monte Carlo by M.H. The variational Monte Carlo was performed by F.F., the pseudofermion functional renormalization group calculations by M.H. Density functional theory calculations were performed by H.O.J. All authors contributed to the manuscript.\\
%\vspace{0.5cm}

{\bf\large Competing interests}\\
The authors declare no competing interests.

\vspace{0.2cm}
{\bf\large Additional information}\\
{\bf Supplementary information} The online version contains supplementary material available at https://doi.org/10.1038/xxx.

\clearpage


\section*{Supplementary material}

# Phase diagram of a distorted kagome antiferromagnet and application to Y-kapellasite – Supplementary Information –

Max Hering,[1,2] Francesco Ferrari,[3] Aleksandar Razpopov,[3] Igor I. Mazin,[4] Roser Valentí,[3] Harald O. Jeschke,[5] and Johannes Reuther[1,2]

[1]Helmholtz-Zentrum Berlin für Materialien und Energie, Hahn-Meitner Platz 1, 14109 Berlin, Germany
[2]Dahlem Center for Complex Quantum Systems and Fachbereich Physik, Freie Universität Berlin, 14195 Berlin, Germany
[3]Institute for Theoretical Physics, Goethe University Frankfurt,
Max-von-Laue-Straße 1, 60438 Frankfurt am Main, Germany
[4]Department of Physics and Astronomy, and Quantum Science and Engineering Center,
George Mason University, Fairfax, VA 22030, USA
[5]Research Institute for Interdisciplinary Science, Okayama University, Okayama 700-8530, Japan
(Dated: November 29, 2021)

## Supplementary Note 1. Finite-size clusters

Most of the numerical calculations of this work are performed on finite-size clusters with periodic boundary conditions. We consider two classes of fully symmetric finite-size clusters, *type-I* and *type-II*, which are defined by two different choices for the boundary conditions. *Type-I* clusters are constructed by imposing periodic boundary conditions along the vectors $\vec{T}_1 = L\vec{a}_1$ and $\vec{T}_2 = L\vec{a}_2$, for a total of $N = 9L^2$ sites. Only the *Type-I* clusters in which $L$ is a multiple of 3 can accommodate the $\vec{Q} = (1/3, 1/3)$ magnetic order. For *type-II* clusters, instead, we take periodic boundary conditions along $\vec{T}_1 = L(\vec{a}_1 - \vec{a}_2)$ and $\vec{T}_2 = L(2\vec{a}_1 + \vec{a}_2)$, which are parallel to the axes of the ideal (undistorted) kagome lattice. *Type-II* clusters contain $N = 27L^2$ sites, and they can always accommodate the $\vec{Q} = (1/3, 1/3)$ order, regardless of the value of $L$. In Supplementary Fig. 1 we show the small finite-size clusters employed for the iterative minimization calculations, i.e. the $L = 2$ *type-I* cluster with $N = 36$ sites and the $L = 1$ *type-II* cluster with $N = 27$ sites.

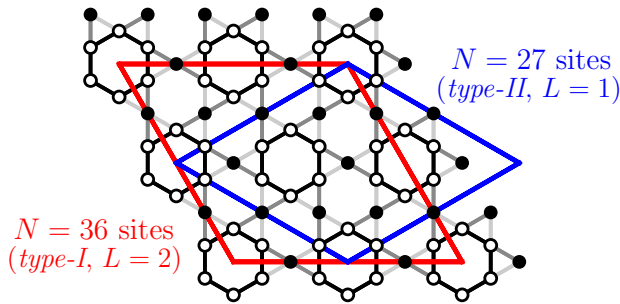

Supplementary Figure 1. Finite-size clusters used for iterative minimization, classical Monte Carlo, and variational Monte Carlo calculations.

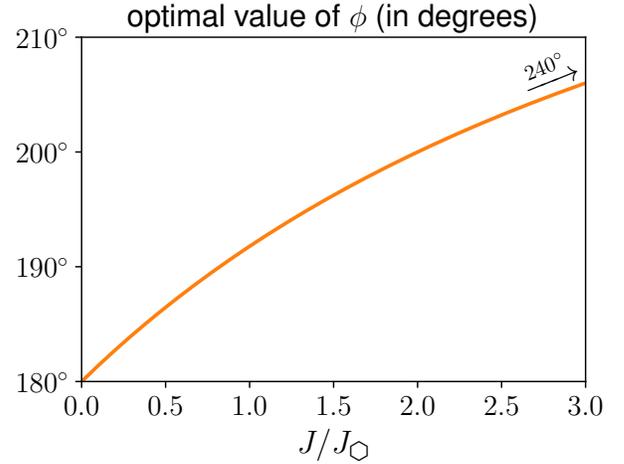

Supplementary Figure 2. Optimal value of the angle $\phi$ for the $\vec{Q} = (1/3, 1/3)$ magnetic order of Fig. 1c of the main text as a function of the $J/J_\bigcirc$ ratio ($J' = 0$).

## Supplementary Note 2. More on the $\vec{Q} = (1/3, 1/3)$ ordered phase

In the limit $J' = 0$, the energy of the classical spin configuration is given by Eq. (7) of the main text. The optimal value of $\phi$ is obtained by the minimization of this energy and its dependency on the ratio $J/J_\bigcirc$ is shown in Supplementary Fig. 2.

Further insight is gained by looking at two extreme limits of the ratio $J/J_\bigcirc$. When $J \ll J_\bigcirc$ (strong hexagons limit), the angle $\phi \to \pi$, namely the spins of $A$ sites form an antiferromagnetic pattern around the $J_\bigcirc$-hexagons. On the other hand, in the opposite limit, $J \gg J_\bigcirc$ (trimer limit), the angle $\phi \to 4\pi/3$ and the spins of sublattice $B$ are antiferromagnetically aligned with respect to the two nearest-neighboring $A$ sites, thus forming antiferromagnetic trimers as shown in the main text, Fig. 1d. Considering the trimers as single effective spins, we note that they form a kagome lattice and they interact through an effective antiferromagnetic coupling (each trimer is connected

| name | $d_{Cu-Cu}$ (Å) | assignment |
|---|---|---|
| $J_1 \equiv J'$ | 3.24978 | inplane 1nn |
| $J_2 \equiv J_\bigcirc$ | 3.36832 | inplane 1nn |
| $J_3 \equiv J$ | 3.37619 | inplane 1nn |
| $J_4$ | 5.67876 | interlayer |
| $J_5$ | 5.70735 | inplane 2nn |
| $J_6$ | 5.76750 | inplane 2nn |
| $J_8$ | 5.86065 | interlayer |

| $U$ (eV) | $J_1$ (K) | $J_2$ (K) | $J_3$ (K) | $J_4$ (K) | $J_5$ (K) | $J_6$ (K) | $J_8$ (K) | $T_{CW}$ (K) |
|---|---|---|---|---|---|---|---|---|
| 4 | 11.3(9) | 161.1(6) | 189.1(9) | -0.5(1.0) | 4.2(6) | 0.5(6) | 2.0(9) | -122 |
| 4.5 | 10.2(7) | 149.3(5) | 174.0(7) | -0.9(8) | 3.8(5) | 0.5(5) | 1.4(7) | -113 |
| 5 | 9.1(6) | 138.7(4) | 160.3(6) | -1.2(7) | 3.4(4) | 0.4(4) | 0.9(6) | -104 |
| **5.23** | **8.7(6)** | **134.2(4)** | **154.4(6)** | **-1.3(7)** | **3.2(4)** | **0.4(4)** | **0.7(6)** | **-100** |
| 5.5 | 8.2(5) | 129.1(3) | 147.8(5) | -1.4(5) | 3.0(3) | 0.4(3) | 0.5(5) | -96 |
| 6 | 7.4(4) | 120.2(3) | 136.4(4) | -1.4(5) | 2.7(3) | 0.3(3) | 0.2(4) | -89 |
| 6.5 | 6.6(3) | 112.0(2) | 125.8(3) | -1.5(4) | 2.3(2) | 0.3(2) | 0.0(3) | -82 |
| 7 | 5.9(3) | 104.4(2) | 116.0(3) | -1.4(3) | 2.0(2) | 0.2(2) | -0.2(3) | -76 |
| 7.5 | 5.2(2) | 97.2(2) | 106.9(2) | -1.4(3) | 1.7(2) | 0.2(2) | -0.3(2) | -70 |
| 8 | 4.5(2) | 90.5(1) | 98.5(2) | -1.3(2) | 1.5(2) | 0.1(2) | -0.4(2) | -65 |

Supplementary Table 1. Exchange couplings of $Y_3Cu_9(OH)_{19}Cl_8$ ($S_{XRD}$ structure), calculated within GGA+U and $6 \times 6 \times 6$ $k$ points. The experimental value $T_{CW} = -100$ K of the Curie Weiss temperature is matched by the line in bold face. Statistical errors are indicated.

| name | $d_{Cu-Cu}$ (Å) | assignment |
|---|---|---|
| $J_1 \equiv J'$ | 3.25310 | inplane 1nn |
| $J_2 \equiv J$ | 3.36068 | inplane 1nn |
| $J_3 \equiv J_\bigcirc$ | 3.37591 | inplane 1nn |
| $J_4$ | 5.68423 | interplane 1nn |
| $J_5$ | 5.68831 | inplane 2nn |
| $J_6$ | 5.76630 | inplane 2nn |
| $J_7$ | 5.77832 | inplane 2nn |

| $U$ (eV) | $J_1$ (K) | $J_2$ (K) | $J_3$ (K) | $J_4$ (K) | $J_5$ (K) | $J_6$ (K) | $J_7$ (K) | $T_{CW}$ (K) |
|---|---|---|---|---|---|---|---|---|
| 4 | 11.2(7) | 151.4(8) | 154.3(5) | -1.0(7) | 5.7(5) | 0.8(5) | 3.3(7) | -108 |
| **4.46** | **10.3(7)** | **139.8(8)** | **143.9(5)** | **-1.3(7)** | **5.2(5)** | **0.8(5)** | **2.6(7)** | **-100** |
| 4.5 | 10.3(6) | 138.8(7) | 143.0(4) | -1.3(6) | 5.2(4) | 0.7(4) | 2.6(6) | -99 |
| 5 | 9.4(5) | 127.4(6) | 132.7(3) | -1.5(5) | 4.6(3) | 0.6(3) | 1.9(5) | -91 |
| 5.5 | 8.5(4) | 117.1(5) | 123.4(3) | -1.6(4) | 4.1(3) | 0.5(3) | 1.4(4) | -84 |
| 6 | 7.7(3) | 107.7(4) | 114.9(2) | -1.6(3) | 3.6(2) | 0.5(2) | 1.1(3) | -78 |
| 6.5 | 6.8(3) | 99.1(3) | 107.0(2) | -1.6(3) | 3.2(2) | 0.4(2) | 0.7(3) | -72 |
| 7 | 6.1(2) | 91.1(3) | 99.6(2) | -1.5(2) | 2.8(2) | 0.3(2) | 0.5(2) | -66 |
| 7.5 | 5.3(2) | 83.8(2) | 92.7(1) | -1.4(2) | 2.4(2) | 0.2(2) | 0.3(2) | -61 |
| 8 | 4.5(2) | 77.0(2) | 86.3(1) | -1.3(2) | 2.1(1) | 0.2(1) | 0.1(2) | -56 |

Supplementary Table 2. Exchange couplings of $Y_3Cu_9(OH)_{19}Cl_8$ ($S_{ND}$ structure), calculated within GGA+U and $6 \times 6 \times 6$ $k$ points. The experimental value $T_{CW} = -100$ K of the Curie Weiss temperature is matched by the line in bold face. Statistical errors are indicated.

to four other trimers by a $J_\bigcirc$ bond). In this limit, the $\vec{Q} = (1/3, 1/3)$ magnetic order of our model reduces to the $\sqrt{3} \times \sqrt{3}$ order [3] of the effective kagome lattice made by the antiferromagnetic trimers.

As discussed in the main text, the linear spin wave dispersion of the $\vec{Q} = (1/3, 1/3)$ order in the *strong hexagon* ($J \ll J_\bigcirc$, Fig.3a) and *trimer* ($J \gg J_\bigcirc$, Fig.3c) limits share similarities with those of the triangular lattice 120° order [1] and of the kagome lattice $\sqrt{3} \times \sqrt{3}$ order [3], respectively. In Supplementary Fig. 3 we show the linear spin wave dispersions for the latter two cases, for comparison.

## Supplementary Note 3. More on the classical spin liquid phase

It is worth mentioning that the $J = J'$ line represents a special case within this region, where coplanar ground states with $\vec{L}_\triangle = 0$ can be analytically defined. Indeed, in this special limit, for each triangle we can write

$$\vec{L}_\triangle = J\vec{S}_b + \vec{S}_a + \vec{S}_{a'},$$ (1)

where $a$ and $a'$ are two neighboring sites of sublattice $A$, $b$ is a site of sublattice $B$ and we have set $J_\bigcirc = 1$ for

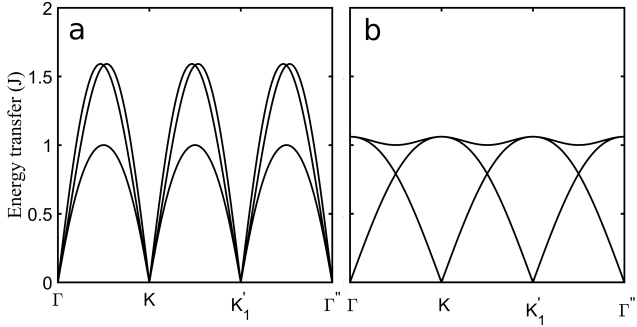

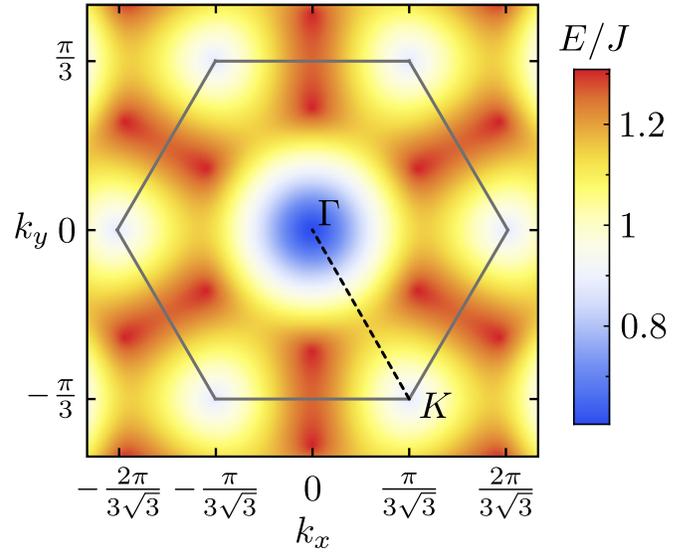

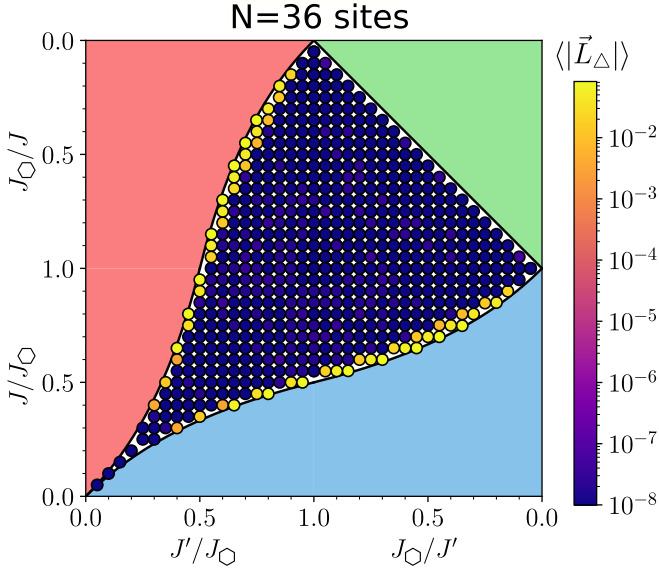

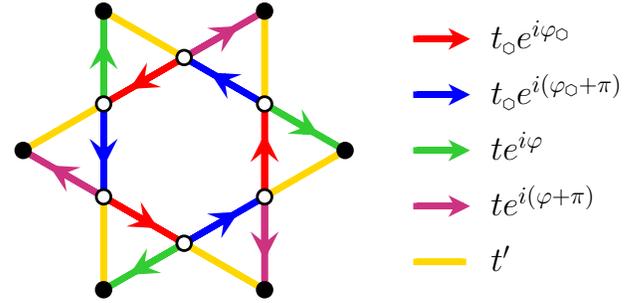

**Supplementary Figure 3.** Linear spin wave dispersions for **a** the $120°$ magnetic order in the antiferromagnetic Heisenberg model on the triangular lattice [1] and **b** the $\sqrt{3} \times \sqrt{3}$ order in the antiferromagnetic Heisenberg model on the kagome lattice [2]. We note that in the latter case the true ground state of the $S = 1/2$ system is not magnetically ordered. A zero-energy dispersionless mode is present in the linear spin-wave spectrum of the $\sqrt{3} \times \sqrt{3}$ order on the kagome lattice [2]. In both panels the spin wave energies are plotted along the $\Gamma - K - K'_1 - \Gamma$ line (see Fig.1b of the main text).

**Supplementary Figure 4.** Value of $\langle |\vec{L}_\triangle| \rangle$ (averaged over triangles) in the optimal classical ground state found by iterative minimization. The results are obtained in the classical spin liquid phase, on the $N = 36$ sites cluster sketched in Supplementary Fig. 1 (*type-I*, $L = 2$, see Supplementary Note 1 for details). Close to the inner boundaries of the spin-liquid phase, we find a number of points where the system is characterized by two distinct non-coplanar degenerate ground states with $\langle |\vec{L}_\triangle| \rangle > 0$. We note that the $N = 36$ sites cluster cannot accommodate the $\vec{Q} = (1/3, 1/3)$ order.

simplicity. Since this expression is valid for all triangles of the system, simple solutions with $\vec{L}_\triangle = 0$, $\forall \triangle$ can be found by focusing on a single unit cell. There, we can set, e.g., $S_b = (1, 0, 0)$ for all sites of sublattice $B$ and we can choose an alternating pattern for the sites of sublattice $A$ forming

**Supplementary Figure 5.** Third band of the coupling matrix [see Eq. (2)] depicted in the first Brillouin zone (gray hexagon). Energies $E$ are indicated relative to the bottom of the lowest band. The dashed line marks a valley in the band structure which may explain the system's unusual thermal fluctuations.

**Supplementary Figure 6.** Optimal hopping pattern of the best variational *Ansatz* [Eq. (5)] for the effective spin model of Y-Kapellasite. The empty (full) dots represent the sites of sublattice $A$ ($B$). The arrows encode the convention for complex hoppings: $i \rightarrow j$ represents the hopping terms $t_{i,j} c_{i,\sigma}^\dagger c_{j,\sigma}$ ($\sigma = \uparrow, \downarrow$). A total of five real parameters is used to parametrize the hopping part of the variational *Ansatz* ($t_\bigcirc, t, t', \varphi_\bigcirc, \varphi$).

the $J_\bigcirc$-hexagons, where we have $S_a = (\cos(\theta), \sin(\theta), 0)$ for odd sites and $S_{a'} = (\cos(\theta), -\sin(\theta), 0)$ for even sites, with $\theta = \arccos(-J/2)$. The same analysis can be repeated for each unit cell separately and coplanar ground states can be constructed, as discussed also in Ref. [4], where the $J = J'$ model is investigated in connection to the physics of Volborthite. We note that this solution is connected to the $\vec{Q} = 0$ coplanar order, since for $J = J' = 2J_\bigcirc$ we get $\theta = \pi$, i.e. the spins inside the $J_\bigcirc$-hexagons are ferromagnetically arranged, and antiparallel to the spins of sublattice $B$.

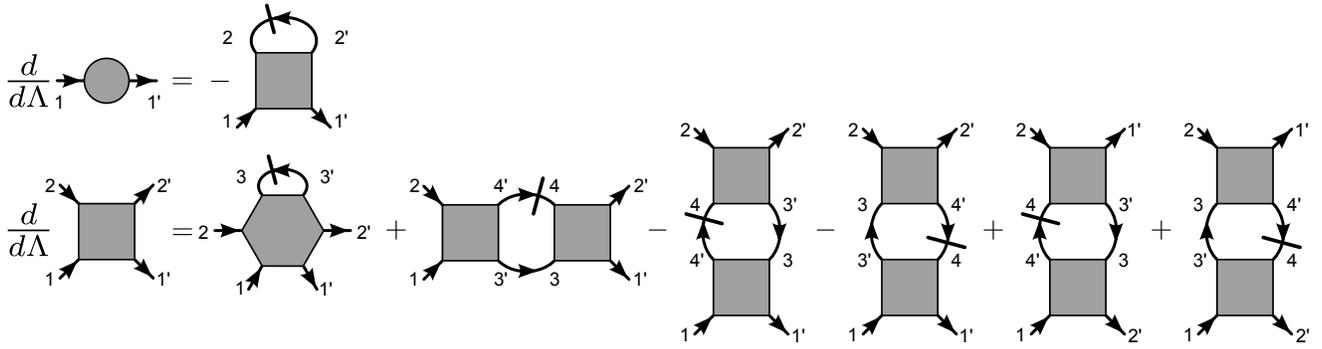

Supplementary Figure 7. Diagrammatic representation of the PFFRG flow equations for $\Sigma^\Lambda$ (circle) and $\Gamma^\Lambda$ (square). Slashed (bare) arrows depict the single-scale propagator $S^\Lambda$ (dressed propagator $G^\Lambda$). Since the two-particle vertex couples to the three-particle vertex (hexagon), the flow equations are truncated at this level. External legs contain one full set of quantum numbers each and internal quantum numbers have to be summed up or integrated.

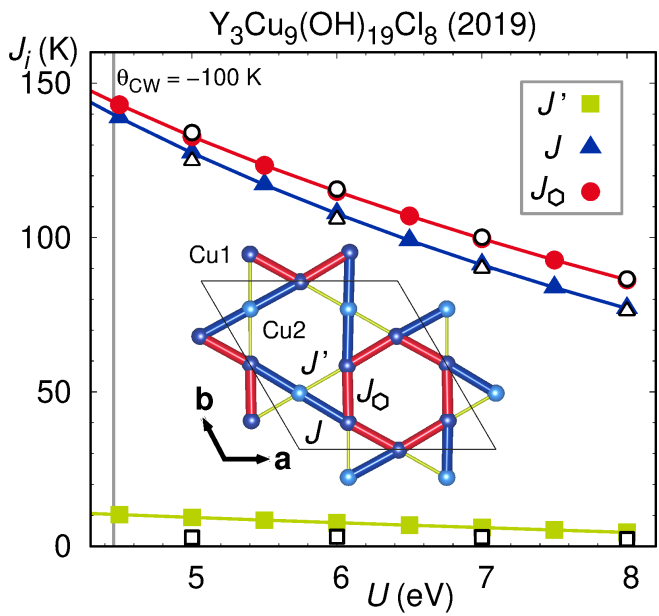

Supplementary Figure 8. Exchange couplings of $Y_3Cu_9(OH)_{19}Cl_8$ ($S_{ND}$ structure) extracted with GGA+U at $J_H = 1$ eV as function of interaction strength $U$. Solid symbols: $P1$ cell, 7 couplings extracted. Empty symbols: $\sqrt{2} \times \sqrt{2} \times 1$ supercell, 24 couplings extracted.

| name | $d_{Cu-Cu}$ (Å) | $J_i$ (K) | assignment |
|---|---|---|---|
| $J_1$ | 3.24978 | 8.4(9) | inplane 1nn Cu1-Cu2 |
| $J_2$ | 3.36832 | 133.1(6) | inplane 1nn Cu1-Cu1 |
| $J_3$ | 3.37619 | 152.0(9) | inplane 1nn Cu1-Cu2 |
| $J_4$ | 5.67876 | 0.3(8) | interlayer |
| $J_5$ | 5.70735 | 1.0(5) | inplane 2nn Cu1-Cu1 |
| $J_6$ | 5.76750 | 0.6(6) | inplane 2nn Cu2-Cu2 |
| $J_7$ | 5.83040 | 3.0(5) | inplane 2nn Cu1-Cu1 |
| $J_8$ | 5.86065 | 0.0(6) | interlayer |
| $J_9$ | 6.49956 | -0.6(5) | inplane 3nn ($2J_1$, Cu1-Cu1) |
| $J_{10}$ | 6.51314 | -0.5(6) | interlayer |
| $J_{11}$ | 6.57249 | -0.5(5) | interlayer |
| $J_{12}$ | 6.61523 | -1.7(7) | inplane 3nn ($J_1+J_2$, Cu1-Cu2) |
| $J_{13}$ | 6.62488 | -0.3(7) | inplane 3nn ($J_1,J_3$ hex., Cu1-Cu2) |
| $J_{14}$ | 6.63443 | -1.6(9) | interlayer |
| $J_{15}$ | 6.66444 | 0.2(7) | interlayer |
| $J_{16}$ | 6.69872 | -0.2(5) | interlayer |
| $J_{17}$ | 6.72699 | 0.4(6) | interlayer |
| $J_{18}$ | 6.73343 | 1.2(5) | inplane 3nn ($J_2$ hex., Cu1-Cu1) |
| $J_{19}$ | 6.74065 | 0.2(8) | inplane 3nn ($J_2+J_3$, Cu1-Cu2) |
| $J_{20}$ | 6.75238 | -0.1(6) | inplane 3nn ($2J_3$, Cu1-Cu1) |
| $J_{21}$ | 8.04603 | 0.7(8) | interlayer |
| $J_{22}$ | 8.09002 | 3.2(5) | interlayer |
| $J_{23}$ | 8.09791 | 0.4(7) | interlayer |
| $J_{24}$ | 8.14162 | 1.0(7) | interlayer |

Supplementary Table 3. Exchange couplings of $Y_3Cu_9(OH)_{19}Cl_8$ ($S_{XRD}$ structure), calculated within GGA+U and $4 \times 4 \times 4$ $k$ points. The couplings are reported at $U = 5.28$ eV; this $U$ value is obtained by demanding that the model matches the experimental Curie-Weiss temperature. Statistical errors are indicated.

## Supplementary Note 4. Analytical approach to explain classical Monte Carlo data

In this Supplementary Note we discuss the origin of the unusual peak shift in the classical Monte Carlo data for the $S_{XRD}$ structure as a function of temperature, as observed in Fig. 5 of the main text. A possible explanation for this behavior comes from the system's coupling matrix in momentum space

$$J_{\rho\sigma}(\vec{k}) = \sum_{\vec{r}_a - \vec{r}_b} e^{i\vec{k}(\vec{R}_a - \vec{R}_b)} J_{a\rho b\sigma} , \quad (2)$$

which forms the basis for various analytical techniques for studying classical spin systems. Here, the sites $i$ are characterized by the two indices $a$ and $\rho$, specifying the unit cell and the sublattice of site $i$, respectively. Similarly, $j \to b, \sigma$ and $J_{ij} \to J_{a\rho b\sigma}$, yielding the $9 \times 9$ matrix $J_{\rho\sigma}(\vec{k})$. Furthermore, $\vec{R}_a$, $\vec{R}_b$ specify the positions of the unit cells $a$ and $b$. Diagonalizing this matrix for the couplings of the $S_{XRD}$ structure one finds that the third band (when counting from the lowest one) is in an energy range from $E \sim 0.6J$ to

$E \sim 1.3J$ relative to the bottom of the lowest band. This intermediate energy range roughly corresponds to the temperature regime at which the peak shift in Monte Carlo is observed, i.e., the third band can be expected to be connected to these unusual thermal fluctuations. Indeed, as shown in Supplementary Fig. 5 the third band exhibits a pronounced valley between the $K$ and $\Gamma$ points (dashed line in Supplementary Fig. 5) such that fluctuations preferably occur along this direction. Note that the magnetic band structure is defined in the first Brillouin zone such that all points $\Gamma'$, $\Gamma''$ are folded back onto $\Gamma$ and, equivalently, $K_1'$, $K_2'$, $K_3'$ all map onto $K$.

**Supplementary Note 5. Details on the variational Monte Carlo calculations**

The variational Monte Carlo (VMC) calculations for the Hamiltonian of Y-kapellasite make use of Gutzwiller-projected fermionic wave functions as variational *Ansätze*. The definition of these states is based on the Abrikosov fermion representation of $S = 1/2$ spins,

$$\vec{S}_i = \frac{1}{2} \sum_{\alpha,\beta} c_{i,\alpha}^\dagger \vec{\sigma}_{\alpha,\beta} c_{i,\beta}. \tag{3}$$

Here $\vec{\sigma} = (\sigma_x, \sigma_y, \sigma_z)$ is a vector of Pauli matrices and $c_{i,\alpha}$ ($c_{i,\alpha}^\dagger$) are fermionic annihilation (creation) operators. Within this formalism, suitable variational states for spin systems can be constructed by applying the Gutzwiller projector, $\mathcal{P}_G = \sum_i (n_{i,\uparrow} - n_{i,\downarrow})^2$, to a fermionic wave function, e.g. a Slater determinant [5]. The Gutzwiller operator enforces the single occupation of each lattice site and projects the fermionic wave function onto the spin Hilbert space. In this work, we consider Gutzwiller-projected Jastrow-Slater variational *Ansätze* of the form

$$|\Psi_0\rangle = \mathcal{J}\mathcal{P}_G |\Phi_0\rangle. \tag{4}$$

Here, $|\Phi_0\rangle$ is the ground state of an auxiliary quadratic Hamiltonian of Abrikosov fermions, named $\mathcal{H}_0$, which includes first-neighbor hopping terms ($t_{i,j}$) and a site-dependent magnetic field of strength $h$:

$$\mathcal{H}_0 = \sum_{\langle i,j \rangle} \sum_\sigma t_{i,j} c_{i,\sigma}^\dagger c_{j,\sigma} + \text{h.c.} + h \sum_i \hat{n}_i^Q \cdot \vec{S}_i. \tag{5}$$

The local orientation of the magnetic field, given by the unit vector $\hat{n}_i^Q$, is chosen such that it induces the optimal single-$\vec{Q}$ classical order in the $(S_x, S_y)$-plane. Since we focus on the exchange couplings derived for the $S_{\text{XRD}}$ structure of Y-kapellasite, our magnetic field corresponds to the best classical order with $\vec{Q} = (1/3, 1/3)$ periodicity. Non-trivial quantum fluctuations are introduced by the hopping terms $t_{i,j}$ of the auxiliary Hamiltonian $\mathcal{H}_0$, and by the application of a long-range spin-spin Jastrow

$$\mathcal{J} = \exp\left[\sum_{i,j} v(i,j) S_i^z S_j^z\right]. \tag{6}$$

on top of the Gutzwiller-projected fermionic state [see Eq. (4)]. The optimal parametrization of the hoppings, which yields the best variational energy for the exchange couplings of the $S_{\text{XRD}}$ structure, is illustrated in Supplementary Fig. 6.

**Supplementary Note 6. Additional DFT results**

In this Supplementary Note we present the full set of Heisenberg exchange coupling constants up to $n^{th}$ neighbors. These values were extracted from spin-polarized *ab initio* DFT calculations by making use of the total energy mapping method as explained in the main text. Supplementary Table 1 shows the results for the $S_{\text{XRD}}$ structure (see Fig. 6 of the main text for the definition of $J_\bigcirc$, $J$ and $J'$) up to $8^{th}$ neighbors, and Supplementary Table 2 shows the results for the $S_{\text{ND}}$ structure up to $7^{th}$ neighbors, where the definitions of $J_\bigcirc$, $J$ and $J'$ are indicated in Supplementary Fig. 8. As explained in the method section of the main text, we perform the energy mapping method by calculating far more total energies than would be minimally required to extract the exchange interactions; for example, for extracting 7 exchange interactions, we calculate 24 energies instead of the minimal number 8. Fitting the Hamiltonian using singular value decomposition yields a statistical error for the exchange couplings which is indicated in Tables 1 to 3. These errors give an indication of the uncertainty of the Hamiltonian parameters calculated for a particular crystal structure; however, the impact of the uncertainties in the crystal structure on the exchange couplings is not included.

In order to check the presence of possible long-ranged coupling constants, we also performed calculations for larger supercells of the $S_{\text{XRD}}$ structure and extracted values of the exchange coupling constants up to $24^{th}$ neighbors, as shown in Table 3. This calculation is computationally very demanding and cannot be repeated for all structures and values of $U$. However, it is an important sanity check because it is the nature of the energy mapping method that a calculation with a small cell yielding few exchange interactions effectively provides information on each coupling with some average of unresolved longer range exchange couplings added to it. By resolving more couplings as shown in Supplementary Table 3, we have verified that there are no important long range exchange couplings that could have introduced errors into the couplings shown in Supplementary Tables 1 and 2 or that would significantly modify the magnetic properties. Comparing the coupling $J_5$ at $U = 5.23$ eV obtained in the smaller cell ($J_5 = 3.2(4)$ K) to that obtained in the large supercell ($J_5 = 1.0(5)$ K), we find that the larger value 3.2 K is not substantiated, and it is reasonable to focus all further analysis on the first three exchange couplings $J_\bigcirc$, $J$ and $J'$.

**Supplementary Note 7.  Comparison of VMC and PFFRG results**

While VMC and PFFRG both find magnetic long range order for Y-kapellasite, the small peak shift and the rather broad susceptibility profile in PFFRG are not seen in VMC, which detects sharp Bragg peaks at the ordering vectors (compare Fig. 7 and Fig. 8 b of the main text). Given that the two techniques are conceptually very different and rely on different approximations, such discrepancies are not surprising. For example, detecting incommensurate magnetic order within VMC is a hard task, because the calculations are restricted to finite-size clusters and thus the vectors of magnetic order are discretized. It is interesting to note that taking a pure hopping *Ansatz* within VMC, i.e. discarding the magnetic field variational parameter (cfr. Supplementary Note 5), we obtain a broad susceptibility profile with shifted peaks, which is very similar to the one provided by PFFRG. However, the corresponding variational state has a considerably higher energy. Possibly, this energy could be reduced if one were able to implement a magnetic field variational parameter for an incommensurate structure, while at the same time maintaining broad susceptibility peaks and thereby reaching better agreement with PFFRG. On the other hand, it is possible that PFFRG and VMC results will resemble each other more closely when promoting the PFFRG towards the aforementioned multi-loop scheme.

---



\end{document}